\documentclass[useAMS,usenatbib]{mn2e}
\usepackage{amsmath}
\usepackage{amssymb}
\usepackage{multirow}
\usepackage{graphicx}
\usepackage{epstopdf}

\def\simlt{\lower.5ex\hbox{$\; \buildrel < \over \sim \;$}}
\def\simgt{\lower.5ex\hbox{$\; \buildrel > \over \sim \;$}}

\newcommand{\bd}{\begin{displaymath}}
\newcommand{\ed}{\end{displaymath}}
\newcommand{\be}{\begin{equation}}
\newcommand{\ee}{\end{equation}}
\newcommand{\beqa}{\begin{eqnarray}}
\newcommand{\eeqa}{\end{eqnarray}}

\newcommand{\Lya}{Ly$\alpha$ }

\title[Parameter Space of the Global 21-cm Signal]
      {Charting the Parameter Space of the Global 21-cm Signal}
      \author[Cohen et al.] {Aviad Cohen$^{1}$\thanks{E-mail:
          aviadc11@gmail.com}, Anastasia Fialkov$^{2}$, Rennan
        Barkana$^{1,3,4,5}$, Matan Lotem$^{1}$ \\ $^{1}$ Raymond and
        Beverly Sackler School of Physics and Astronomy, Tel Aviv
        University, Tel Aviv 69978, Israel\\ $^{2}$ Harvard-Smithsonian Center for Astrophysics, Institute for Theory and Computation, 60 Garden Street, Cambridge, MA 02138, USA
        \\ $^{3}$ Sorbonne Universit\'{e}s,
        Institut Lagrange de Paris (ILP), Institut d'Astrophysique de
        Paris, UPMC Univ Paris 06/CNRS\\ $^{4}$ Department of
        Astrophysics, University of Oxford, Denys Wilkinson Building,
        Keble Road, Oxford OX1 3RH, UK\\ $^5$ Perimeter Institute for
        Theoretical Physics, 31 Caroline St N., Waterloo, ON N2L 2Y5,
        Canada
  }

\begin{document}
\pagerange{\pageref{firstpage}--\pageref{lastpage}} \pubyear{2016}
\maketitle

\label{firstpage}

\begin{abstract} 
The early star-forming Universe is still poorly constrained, with the
properties of high-redshift stars, the first heating sources, and
reionization highly uncertain. This leaves observers planning 21-cm
experiments with little theoretical guidance. In this work we explore
the possible range of high-redshift parameters including the star
formation efficiency and the minimal mass of star-forming halos; the
efficiency, spectral energy distribution, and redshift evolution of
the first X-ray sources; and the history of reionization. These
parameters are only weakly constrained by available observations,
mainly the optical depth to the cosmic microwave background. We use
realistic semi-numerical simulations to produce the global 21-cm
signal over the redshift range $z = 6-40$ for each of 193 different
combinations of the astrophysical parameters spanning the allowed
range. We show that the expected signal fills a large parameter space,
but with a fixed general shape for the global 21-cm curve. Even with
our wide selection of models we still find clear correlations between
the key features of the global 21-cm signal and underlying
astrophysical properties of the high redshift Universe, namely the
Ly$\alpha$ intensity, the X-ray heating rate, and the production rate
of ionizing photons. These correlations can be used to directly link
future measurements of the global 21-cm signal to astrophysical
quantities in a mostly model-independent way.  We identify additional
correlations that can be used as consistency checks.

\end{abstract}


\section{Introduction}
\label{Sec:Intro}

Some of the most exciting epochs in cosmic history, including the
cosmic dark ages, the formation of the first radiative sources (cosmic
dawn), and the onset of the epoch of reionization during
which the entire Universe became ionized, are currently inaccessible
observationally. Our theoretical understanding of galaxy formation
gives us significant guidance, but this is limited by astrophysical
uncertainties \citep{Barkana:2016}. A major focus are three cosmic
events expected at early times \citep{Madau:1997}: cosmic reionization
(known to have occurred given the highly ionized Universe at present
\citep{GP}), cosmic heating (likely by X-rays), and \Lya coupling (an
event specific to 21-cm cosmology).

In the hierarchical picture of structure formation, halos grew
gradually during the dark ages, assembling mass via gravitational
interactions. Massive enough halos were able to retain gas which could
radiatively cool, condense and form stars, with the first stellar
objects forming at $z\sim 65$ \citep{Naoz:2006, Fialkov:2012}. The
minimal mass of halos within which stars can form, M$_{\textrm{min}}$,
depends on the chemical composition of the gas, and in the pristine
conditions at high redshifts, two cooling channels dominate: (1)
radiative cooling of molecular hydrogen happens in the smallest halos,
with mass above $10^5\,$M$_{\odot}$ \citep[e.g.,][]{Tegmark:1997,
  Bromm:2002, Yoshida:2003}, and (2) radiative cooling of atomic
hydrogen takes place in halos with mass above $10^7\,$M$_\odot$
\citep[e.g.,][]{Barkana:2001}. Star formation in small halos is a
vulnerable process and is believed to be affected by several feedback
mechanisms which can either boost or suppress the formation of the
next generation of stars. One of the mechanisms discussed in the
literature is the Lyman-Werner (LW) feedback. UV radiation in the LW
band emitted by the first stars can dissociate hydrogen molecules
\citep{Haiman:1997}, depleting the reservoirs of gas available for the
formation of the future stars \citep[however, the efficiency of the LW
  feedback is poorly understood,
  e.g.,][]{Visbal:2014,Schauer:2015}. Because LW photons reach up to
$\sim 100$ comoving Mpc, this feedback is not local and star formation
activity at one site can potentially sterilize halos over a large
cosmological volume. Another possible feedback mechanism is the
stellar feedback from supernova explosions which can expel gas from
the halo, effectively increasing the minimum cooling mass well above
the atomic cooling threshold \citep{Wyithe:2013}. An additional
feedback mechanism that can affect star formation is the photoheating
feedback which becomes efficient once the intergalactic gas is
photoheated above $10^4$ K by ionizing radiation emitted by stars;
this gas stops accreting onto halos below $\sim 10^8-10^9$ M$_\odot$,
thus quenching subsequent star formation within them. Because heavy
halos are rare at high redshifts, LW, supernova and photoheating feedbacks
can, when they are effective, delay major cosmological events such as
the heating of the intergalactic gas and reionization. Finally, there
is a possibility that light halos (below the atomic cooling mass) can
continue to contribute to star formation even in the presence of LW
radiation, via the metal-line cooling channel. Because metal-line
cooling is more efficient than molecular cooling, this channel can
dominate star formation in small halos once the gas is enriched by the
first supernovae explosions. However, the possibility of star
formation via metal cooling in the early universe and its contribution
to the total star formation is highly uncertain
\citep[e.g.,][]{Jeon:2014,Wise:2014,O'Shea:2015,Cohen:2016}.
 
The fraction of gas that is converted into stars (the star
  formation efficiency, hereafter SFE) is another unknown and can be
of order a few tens of percent or lower depending on the halo mass,
redshift and dominant feedback mechanisms. Observations at low
redshifts show that the star formation efficiency is a few percent in
massive halos \citep{Tinker:2016}, while isolated dwarf galaxies show
a very low SFE of order $\sim 0.1-0.01$ \%
\citep{Read:2016}. Simulations of high-redshift stellar activity
present a large scatter of values for SFE, especially for small halos
which likely dominated the early Universe \citep[e.g.,][]{Wise:2014,
  O'Shea:2015, Xu:2016}. Matching the observed luminosity function to
the expected number of halos at $z\gtrsim 6$ shows that the peak value
of the SFE is 30\% for halos of M$_\textrm{h}\sim 10^{11}-10^{12}$
M$_\odot$, dropping to SFE$\sim 10\%$ at the low mass
M$_\textrm{h}\sim 2\times 10^{10}$ M$_\odot$ and high mass
M$_\textrm{h}\sim 3\times 10^{13}$ M$_\odot$ limits
\citep{Behroozi:2015, Mirocha:2016, Mason:2015, Mashian:2016,
  Sun:2016}.

As noted above, the formation of the first luminous objects had a
dramatic effect on the Universe, completely changing the
environment. The first astronomical objects emitted UV and X-ray
radiation which heated and ionized the gas while supernova explosions
enriched the primordial gas with metals leading to the formation of
the next generation of stars. Stars are believed to have been the main
origin of UV photons which reionized the neutral intergalactic medium
(IGM), resulting in the total cosmic microwave background (CMB)
optical depth of $\tau \sim 0.055\pm0.009$
\citep{Planck:2016b}. However, the origin of the first heating
sources, which raised the temperature of the IGM above that of the
CMB, is still debatable. The most plausible heating radiation is
X-rays, which can travel far even in a neutral Universe. The X-ray
efficiency of the sources as well as their spectral energy
distribution (SED) remain very poorly constrained. Several different
candidates have been proposed in the literature including X-ray
binaries (XRBs) \citep{Mirabel:2011}, mini-quasars \citep{Madau:2004},
hot gas in the first galaxies, and hard X-rays produced via inverse
Compton scattering of the CMB off electrons accelerated by supernovae
\citep{Oh:2001}. Finally, there are more exotic possibilities such as
dark matter annihilation \citep{Cirelli:2009}. Out of the plethora of
candidates, XRBs (which have a hard SED which peaks around $1-3$~keV)
are likely to be the dominant source of cosmic heating at $z\gtrsim 6$
\citep{Mirabel:2011,Fragos:2013}. The hard spectrum has a major effect
on 21-cm cosmology, substantially delaying cosmic heating and
decreasing the amplitude of 21-cm fluctuations from heating
\citep{Fialkov:2014b}. Extrapolations of recent observations to high
redshift continue to support such a scenario
\citep{Madau:2016,Mirocha:2016}. However, direct observational
constraints on the X-ray efficiency of the first sources are rather
weak. Upper limits on the heating efficiency come from the soft
unresolved cosmic X-ray background \citep{Fialkov:2016} and lower
limits are given by the observed upper limits on the 21-cm power
spectrum \citep{Ali:2015, Pober:2015, Fialkov:2016}.

The most promising tool to explore the early universe is the
redshifted 21-cm signal of neutral hydrogen. It is strongly affected
by astrophysics and cosmology, and, thus, is believed to be an
excellent probe of processes that took place at high redshifts. In
particular, the first stars also are expected to have emitted
Ly$\alpha$ photons (plus higher energy photons that redshifted down to
Ly$\alpha$), which coupled the 21-cm line (in terms of the relative
abundance of its ground and excited states) to the kinetic gas
temperature, leading to a strong, potentially observable, 21-cm signal
\citep{Madau:1997}, which otherwise would have faded away by $z \sim
30$.

The currently unexplored parameter space of the early universe leaves
a large window within which the 21-cm signal may fall, making it
difficult to predict its shape and guide current and future radio
telescopes. The signal has not been detected
yet \footnote{\citet{Bernardi:2016} used a Bayesian method with a
  simplified Gaussian model for the absorption feature (see
  Section~\ref{Sec:Methods}) to constrain the global signal using
  early LEDA observations.} and only upper limits have been placed on
its power spectrum at redshift $z<10$
\citep{Ali:2015,Pober:2015,Ewall-Wice:2016}; however, many current and
future observations aim to detect and measure the signal out to
$z\sim35$. Experiments such as the Experiment to Detect the Global EoR
Step \citep[EDGES,][]{Bowman:2010}, the Shaped Antenna measurement of
the background RAdio Spectrum \citep[SARAS,][]{Patra:2013}, the Large
Aperture Experiment to Detect the Dark Age
\citep[LEDA,][]{Bernardi:2015} and the Dark Ages Radio Explorer
\citep[DARE,][]{Burns:2012}, are trying to measure the global signal,
while the Low Frequency Array \citep[LOFAR,][]{van Haarlem:2013}, the
Murchison Wide-field Array
\citep[MWA,][]{Bowman:2013,Ewall-Wice:2016}, the Precision Array to
Probe the Epoch of Reionization \citep[PAPER,][]{Ali:2015}, the
Hydrogen Epoch of Reionization Array
\citep[HERA,][]{Pober:2014,DeBoer:2016}, and the Square Kilometer
Array \citep[SKA,][]{Koopmans:2015} are aiming to measure the power
spectrum.

Our goal in this paper is to explore the full parameter space of the
global 21-cm signal resulting from the uncertainties in the
astrophysical parameters of the high-redshift universe. Other recent
work has focused on extrapolating low-redshift observations of
galaxies to high redshift \citep{Madau:2016,Mirocha:2016}, but we
adopt a more flexible approach. While it will be interesting to use
observations to find out if such extrapolations are accurate,
a~priori, this cannot be assumed. Compared to current observations
(which are mostly at relatively low redshift), conditions are very
different at redshift 20, e.g., in terms of the CMB temperature, the
cosmic and virial halo densities (of both the dark matter and gas),
the typical mass of galactic halos, and halo merger histories.  Thus,
the astrophysical properties of early galaxies could be quite
different from those suggested by extrapolations of observed galaxies,
and it is important to keep an open mind until direct observational
evidence becomes available.

In what follows, as we lay out the large parameter space
  possible for the global 21-cm signal, we try to characterize the
  properties of this signal and find relations between the shape of
  the global signal and the astrophysical parameters at high
  redshifts.  \cite{Mirocha:2013} previously addressed parameter
  reconstruction using a physical model for the global signal. In this
  \citep[as well as the follow-up works by][where the authors study
    how well current and near-future experiments could constrain the
    four parameters of their model using the measurements of the
    signal's three key points and taking into account the foreground
    and the noise]{Mirocha:2015, Harker:2016}, the authors used
  analytical formulas or simple models that account only for the mean
  evolution of the Universe. In contrast, our more realistic
  simulations include spatial fluctuations in star formation and take
  into account the finite effective horizons of the radiative
  backgrounds, spatially inhomogeneous feedback processes, and time
  delay effects. We also capture a wider parameter space, as our code
  includes the possibility of having substantial star formation in
  halos below the atomic cooling threshold, in which case
  spatially-inhomogeneous processes such as the streaming velocity and
  LW feedback play a key role (and are included in our 21-cm code but
  not in others).

This paper is organized as follows: In Section~\ref{Sec:Methods} we
briefly discuss the general properties of the 21-cm signal as well as
our numerical methods. We present and discuss our specific choice of
the parameters and their ranges in Section~\ref{Sec:Param}, and show
the resulting parameter space spun by the 21-cm signal in
Section~\ref{Sec:results}. Finally, we summarize our results and
discuss our conclusions in Section~\ref{Sec:sum}.

\section{Simulated 21-cm Signal}
\label{Sec:Methods}

In order to explore the parameter space of the early universe and
produce a library of possible global 21-cm signals in the redshift
range $z=6-40$ we use a semi-numerical approach \citep{Mesinger:2011,
  Visbal:2012, Fialkov:2014b}. Our code is a combination of numerical
simulation and analytical calculations and has enough flexibility to
explore the large dynamical range of astrophysical parameters. We
simulate large cosmological volumes of the universe ($384^3$ Mpc$^3$;
all distances comoving unless indicated otherwise) with a 3 Mpc
resolution, and the outcome of the simulation is the resulting
inhomogeneous 21-cm signal which for our purposes in this paper we
average over the box. In addition, inhomogeneous backgrounds of X-ray,
Ly$\alpha$, LW and ionizing radiation at every redshift are
computed. In our simulation, the statistical-generated initial
conditions for structure formation, i.e., the density field and the
supersonic relative velocity between dark matter and baryons
\citep{Tseliakhovich:2010, Tseliakhovich:2011, Visbal:2012}, are
linearly evolved from recombination to lower redshifts. Using the
values of large-scale density and velocity in each cell, we apply the
extended Press-Schechter formalism \citep{Barkana:2004}, as modified
by the large scale density fluctuations and the supersonic relative
velocities, to calculate the local fraction of gas in collapsed
structures in each pixel and at each redshift. We then populate each
pixel with stars given the star formation efficiency, as described in
Section \ref{Sec:Param}. To calculate the intensities of the
  various radiative backgrounds we use the star formation rate (SFR),
  which is determined by the time derivative of the collapsed fraction
  and the SFE.  We use the standard spectra of population II stars
  from \citet{Barkana:2005b} (based on \citet{Leitherer:1999}) to
  determine the spectrum and intensity of Ly$\alpha$ and LW photons,
the strong LW feedback from \citet{Fialkov:2013} (when LW feedback is
applied), and the standard cosmological parameters
\citep{Planck:2014}. Star formation is also subject to the
photoheating feedback \citep{Sobacchi:2013, Cohen:2016}.

The observed cosmic mean 21-cm brightness temperature relative to the
CMB can be expressed as
\citep{Madau:1997,Furlanettoetal:2006,Barkana:2016}:
\begin{equation}
\label{eq:signal}
T_{\rm b}=26.8\, x_{\rm HI}\left( \frac{1+z}{10}\right) ^{1/2}
\left(1+\delta \right) \left[1-\frac{T_{\rm CMB}}{T_S} \right] \rm mK,
\end{equation}
where $x_{\rm HI}$ is the neutral hydrogen fraction, $\delta$ is the
matter overdensity, $T_{\rm CMB}$ is the CMB temperature and $T_S$ is
the spin temperature, which can be expressed as:
\begin{equation}
\label{eq:Ts}
T_S^{-1}=\frac{T_{\rm CMB}^{-1}+x_c T_{\rm gas}^{-1}+x_\alpha
  T_c^{-1}}{1+x_c+x_\alpha} \rm.
\end{equation}
Here $T_{\rm gas}$ is the (kinetic) gas temperature, $T_c$ is the
effective (color) Ly$\alpha$ temperature which is very close to
$T_{\rm gas}$, and $x_c$ and $x_\alpha$ are the coupling coefficients
for collisions and Ly$\alpha$ scattering, respectively.  In
Eq.~(\ref{eq:signal}) we neglect the peculiar velocity term since in
the global signal it averages out to linear order and adds only a tiny
correction \citep{Bharadwaj:2004,Barkana:2005a}.

\begin{figure}
	\centering
	\includegraphics[width=3.2in]{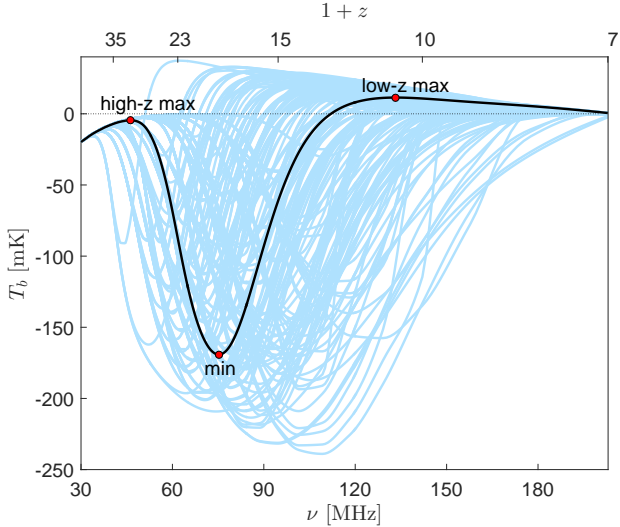}
	\caption{The 21-cm global signal as a function of redshift for
          our standard case (black line), with red points marking the
          three turning points (from left to right: the high-$z$
          maximum, the minimum, and the low-$z$ maximum). Light-blue
          lines show the entire set of realizations of the 21-cm
          signal for the 193 different astrophysical models discussed
          in this paper and summarized in
          Table~\ref{table:variations}. The full list of models
          appears in Appendix~\ref{appA}.  }
	\label{fig:T21cmStacked}
\end{figure}

A typical dependence of the sky-averaged signal (``the 21-cm global
signal'') on frequency is shown in Figure~\ref{fig:T21cmStacked}
(black line, our standard case as will be explained below). Its
characteristic structure of peaks and troughs encodes information
about global cosmic events \citep{Furlanetto:2006}. At early times
$z\gtrsim 40$ collisions between neutral hydrogen atoms and each other
(and with other species) drive $T_S\rightarrow T_{\rm gas}$, and the
signal is seen in absorption, because in the absence of heating
sources $T_{\rm gas}<T_{\rm CMB}$. As the universe expands and cools,
collisions become rare and hydrogen atoms are driven toward thermal
equilibrium with the CMB; thus, the 21-cm brightness temperature
(measured relative to the CMB temperature) decreases in magnitude and
the signal nearly vanishes. When the luminous sources turn on
significantly, the signal reaches a local maximum (still being
observed in absorption). We refer to this point as the ``high-$z$
maximum point'', and it happens at $z\sim30$ for our standard case
(and is equivalent to the turning point B in the nomenclature of
\citet{Mirocha:2013}). As the first sources of Ly$\alpha$ photons turn
on, Ly$\alpha$ radiation begins to drive $T_S$ to $T_c\sim T_{\rm
  gas}$ via the Wouthuysen-Field effect \citep{Wouthuysen:1952,
  Field:1958}. This transition normally occurs before many X-ray
sources turn on and thus the signal is seen in absorption. Once \Lya
coupling approaches saturation, and X-ray sources turn on
significantly, the signal reaches its local minimum value, $\sim -170$
mK in our standard case. We refer to this point as the ``minimum
point'' (equivalent to turning point C of \citet{Mirocha:2013}), and
it occurs at $z\sim18$ for our standard case. At this time the
population of heating sources steadily increases and the contrast
between $T_{\rm gas}$ and $T_{\rm CMB}$ decreases. As the X-ray sources
heat the gas, if the gas temperature rises above that of the CMB, the
21-cm signal is seen in emission. As soon as reionization starts, the
fraction of neutral gas decreases, thus decreasing the amplitude of
the 21-cm signal. In most models (including our standard case), X-ray
sources manage to heat the gas above the CMB temperature, and once
heating saturates (i.e., reaches $T_{\rm gas} \gg T_{\rm CMB}$), another
local maximum, the ``low-$z$ maximum point'' ($z\sim10$, turning point
D of \citet{Mirocha:2013}), is formed. The advance of reionization and
decrease of redshift decrease the signal (see Eq.~\ref{eq:signal}),
until the end of reionization makes $x_{\rm HI}=0$ and the signal vanishes
(we neglect the small fraction of neutral gas left over inside
galaxies). The three key points (within the relevant redshift range)
in the evolution of the signal are marked with red dots in
Figure~\ref{fig:T21cmStacked}.

\renewcommand{\thefootnote}{\fnsymbol{footnote}}

\begin{table*}
\begin{center}
	\begin{tabular}{llllll}
		\hline \begin{tabular}[c]{@{}l@{}}{\bf
                    Category} \end{tabular}
                & \begin{tabular}[c]{@{}l@{}}{\bf Cooling
                      Channel} \end{tabular}
                & \begin{tabular}[c]{@{}l@{}}${\bf f_*}$\end{tabular}
                & \begin{tabular}[c]{@{}l@{}}${\bf f_X}$\end{tabular}
                & \begin{tabular}[c]{@{}l@{}}{\bf SED} \end{tabular}
                & \begin{tabular}[c]{@{}l@{}}\boldmath$\tau$\end{tabular}
                \\ \hline \begin{tabular}[c]{@{}l@{}}Standard
                  Case\\ (1 model)\end{tabular}
                & \begin{tabular}[c]{@{}l@{}}Atomic
                    cooling\end{tabular} &
		\begin{tabular}[c]{@{}l@{}}0.05\end{tabular}                                                                   & \begin{tabular}[c]{@{}l@{}}1\end{tabular}                                                                &  \begin{tabular}[c]{@{}l@{}}Hard SED\end{tabular}                                                                                         & \begin{tabular}[c]{@{}l@{}}0.066\end{tabular}    
		                                               \\
		                                               \hline
		\begin{tabular}[c]{@{}l@{}}Small Variations\\ (32 models)\end{tabular} & \begin{tabular}[c]{@{}l@{}}Molecular cooling \\ Massive cooling\end{tabular}& \begin{tabular}[c]{@{}l@{}}$0.05\times\sqrt{10}$\\ $0.05/\sqrt{10}$\end{tabular}    & \begin{tabular}[c]{@{}l@{}}$\sqrt{10}$\\ $1/\sqrt{10}$\end{tabular} & \begin{tabular}[c]{@{}l@{}}Hard SED \& Mini-quasars\\ Soft SED \& Mini-quasars\end{tabular} & \begin{tabular}[c]{@{}l@{}}0.066 \\ 0.082\end{tabular} \\
		\hline
		\begin{tabular}[c]{@{}l@{}}Large Variations\\ (20 models)\end{tabular}  & \begin{tabular}[c]{@{}l@{}}Metal cooling\\ Super-massive cooling\end{tabular}& \begin{tabular}[c]{@{}l@{}}0.5\\ 0.005\end{tabular}                          & \begin{tabular}[c]{@{}l@{}}0.1\\ 10\end{tabular}                & \begin{tabular}[c]{@{}l@{}}Soft SED\\ Mini-quasars\end{tabular}                                 & \begin{tabular}[c]{@{}l@{}}0.066 \\ 0.098\end{tabular}\\
		\hline
		\begin{tabular}[c]{@{}l@{}}Space filler\\ (106 models)\end{tabular}  & \begin{tabular}[c]{@{}l@{}}Molecular cooling\\ Atomic cooling \\ Massive cooling \end{tabular}& \begin{tabular}[c]{@{}l@{}}0.005\\0.05\\ 0.5\end{tabular}                          & \begin{tabular}[c]{@{}l@{}}0.1\\1\\ 8\end{tabular}                & \begin{tabular}[c]{@{}l@{}}Soft SED\\ Hard SED\end{tabular}                                 & \begin{tabular}[c]{@{}l@{}}0.066 \\ 0.082\end{tabular}\\
		\hline
		\begin{tabular}[c]{@{}l@{}}\citet{Fialkov:2016}\\ (22 models)\end{tabular} & \begin{tabular}[c]{@{}l@{}}Atomic cooling\\ Massive cooling\end{tabular}& \begin{tabular}[c]{@{}l@{}}0.05 \end{tabular}                          & \begin{tabular}[c]{@{}l@{}}lower limit\\1\\ upper limit\end{tabular}                & \begin{tabular}[c]{@{}l@{}}Soft SED\\Hard SED\\ Mini-quasars\end{tabular}                                 & \begin{tabular}[c]{@{}l@{}}$0.06-0.11$\end{tabular}\\
		\hline
	\end{tabular}
		\caption{Summary of the considered models.  The name
                  of the category of models appears in the first
                  column, for reference. We vary the cooling channel
                  (column 2), star formation efficiency ($f_*$, column
                  3), X-ray efficiency of X-ray sources ($f_X$, column
                  4), spectral energy distribution of X-ray sources
                  (SED, column 5), and the total CMB optical depth
                  ($\tau$, column 6), taking various combinations of
                  the various parameters within each category. For
                  each cooling channel, we use all the possible
                  combinations of the parameters $f_*$, $f_X$, SED and
                  $\tau$ that are listed in the same category. Note
                  that some of the parameter combinations are ruled
                  out by PAPER measurements and/or they produce only
                  very large $\tau>0.098$ and thus fail our
                  normalization criterion. These models are included
                  here but are excluded from our results. Also, in the
                  \citet{Fialkov:2016} category, each case has a
                  different lower and upper limit on $f_X$ (see
                  Section~\ref{Sec:Xray}). A complete listing of the
                  details of all the included models is given in
                  Appendix~\ref{appA}.}
		\label{table:variations}
	\end{center}
\end{table*}
\renewcommand*{\thefootnote}{\arabic{footnote}}

\section{Model Details and Parameter Ranges}
\label{Sec:Param}
As discussed in the introduction, high-redshift astrophysical
parameters such as the star formation efficiency, X-ray efficiency and
SED, and feedback mechanisms, are poorly constrained. However, they
have a strong impact on the 21-cm signal, affecting the location and
amplitude of the main features of the global signal. To survey
possible realizations we ran our simulation code for 193 different
sets of astrophysical parameters chosen from the ranges described
below, and analyzed the properties of the global 21-cm signal in each
case. Table~\ref{table:variations} presents a brief summary of the
considered models, while the full list of models and their parameter
values is in Appendix~\ref{appA}.

\subsection{Star Formation Efficiency and Cooling Channel}

The star formation efficiency is believed to vary with halo mass,
redshift and metallicity of the gas, and it also depends on feedback
mechanisms. It strongly affects the shape of the global 21-cm signal
by influencing the amount of radiation produced by stars. For
otherwise identical astrophysical parameters, a higher SFE implies an
earlier onset of Ly$\alpha$ coupling, and a faster build-up of X-ray
and ionizing radiation backgrounds. The 21-cm absorption feature is
shallower than in the case of low SFE, because cosmic heating turns on
earlier and the gas does not have as much time to cool down. As a
function of the SFE, all the key points of the global 21-cm signal are
shifted to lower (higher) frequencies in the case of a higher (lower)
SFE.

The high-redshift value of the SFE in the small halos where the first
population of stars formed is highly unconstrained, due to the lack of
direct observations. Existing simulations suggest relatively low
values of the SFE, but show a large scatter \citep[e.g.,][]{Wise:2014,
  O'Shea:2015, Xu:2016}. Based on the low-redshift observations,
typical values of the SFE used in the literature are a few percent
\citep[e.g.,][]{Furlanettoetal:2006,Mesinger:2016,Fialkov:2016}. However,
not only is the typical value of the SFE uncertain, its dependence on
halo mass at the low-mass end is unclear as well \citep{Behroozi:2015,
  Mirocha:2016, Mason:2015, Mashian:2016, Sun:2016}. Therefore, as in
\citet{Cohen:2016}, we consider two possibilities for the SFE-M$_{\rm
  h}$ dependence: a sharp low-mass cutoff
\begin{equation}
f_*(M) = \left\{ \begin{array}{ll}
f_* & M_{\rm min}<M\ ,\\
0 & \mbox{otherwise ,}\end{array} \right. 
\label{eq:fstarSharp} 
\end{equation}
and a gradual low-mass cutoff \citep{Machacek:2001,Fialkov:2013}:
\begin{equation}
f_*(M) = \left\{ \begin{array}{ll}
f_* & M_{\rm atomic} < M\ ,\\
f_*\frac{\log\left( M/M_{\rm min}\right) }{\log\left( M_{\rm atomic}/M_{\rm min}\right) } & 
M_{\rm min} < M < M_{\rm atomic}\ ,\\
0 & \mbox{otherwise ,}\end{array} \right.
\label{eq:fstarGrad}
\end{equation}
where $M_{\rm min}$ is the minimum halo mass for star formation,
$M_{\rm atomic}$ is the minimum halo mass for atomic cooling, and
$f_*$ is a (constant) parameter that stands for the SFE at the
high-mass end. Here we use $f_*=5\%$ as our standard value while also
adopting the values 0.5\% and 50\% when exploring the parameter
space. Note that in both Eqs.~(\ref{eq:fstarSharp}) and
(\ref{eq:fstarGrad}), the SFE-M$_{\rm h}$ dependence is flat for halos
above the atomic cooling threshold, though we also vary the minimum
halo mass widely, including up to values well above this threshold.

The parameter $M_{\rm min}$ is determined either by the cooling
channel through which stars can form, or by feedback. In the
hierarchical picture of structure formation, low-mass halos form at
higher redshifts and are more numerous than high-mass halos at early
times. Therefore, in the cases with lower $M_{\rm min}$, stars form
earlier, leading to an earlier Ly$\alpha$ coupling of the 21-cm signal
to $T_{\rm gas}$ and shifting the location of the high-$z$ maximum point
to higher redshift.

As described in Section~\ref{Sec:Intro}, high-redshift star formation
could happen via several different channels, with each cooling
mechanism having a different minimum cooling mass, which evolves with
redshift. For simplicity, we use the minimum virial circular velocity,
$V_c$, instead of the minimal cooling mass throughout this paper,
since $V_c$ is less strongly dependent on redshift
\citep{Barkana:2016}. To probe different cooling and feedback
mechanisms, we consider here five different scenarios:
\begin{itemize}
	\item Molecular cooling halos: in this case stars can form in
          halos with masses down to the cooling mass of molecular
          hydrogen, i.e., $V_c=4.2$ km s$^{-1}$ which corresponds to
          $M_{\rm vir}\approx7\times10^5M_\odot$ at $z=20$ if LW
          feedback is turned off. In all cases with molecular cooling
          we include LW feedback and star formation efficiency with
          the gradual low-mass cutoff (Eq.~\ref{eq:fstarGrad}).
	\item Metal cooling halos: same as molecular cooling halos but
          without LW, which does not significantly affect the cooling
          of metal-rich gas, and with star formation efficiency with
          the sharp low-mass cutoff (Eq.~\ref{eq:fstarSharp}), to
          obtain the maximal effect (this is the ``Maximal'' case from
          \citet{Cohen:2016}). We note that all cases with small halos
          are significantly affected by the supersonic streaming
          velocity, which significantly and inhomogeneously suppresses
          star formation, while halos above the atomic cooling are
          only weakly affected by it \citep{Tseliakhovich:2010,
            Tseliakhovich:2011, Visbal:2012}.
	\item Atomic cooling halos: stars form in halos with masses
          down to the cooling threshold of atomic hydrogen, i.e., with
          $V_c=16.5$ km s$^{-1}$ (corresponds to $M_{\rm
            vir}\approx3\times10^7M_\odot$ at $z=20$).
	\item Massive halos: stars formation occurs in halos with
          masses down to $10\times M_{\rm atomic}$ which corresponds
          to $V_c=35.5$ km s$^{-1}$.
	\item Super-massive halos: stars form in halos with masses
          down to $100\times M_{\rm atomic}$ ($V_c=76.5$ km
          s$^{-1}$). This or the previous case might correspond to
          strong supernova feedback that expels all the gas out of
          low-mass halos.
\end{itemize}

\subsection{X-ray SED and Normalization}
\label{Sec:Xray}
X-ray heating strongly affects the expected global 21-cm signal by
affecting the depth and the location of the absorption trough, i.e.,
the minimum point, as well as the subsequent rise towards an emission
signal. More efficient X-ray sources imply a shallower absorption
trough with its location shifted to higher redshift, plus a higher
emission signal at the low-$z$ maximum point. On the other hand, weaker
heating results in a deeper trough shifted to lower redshift, with a
suppressed or vanishing emission signal. The energy that goes into
heating of the IGM depends on the total X-ray energy emitted in the
band $\sim 0.2-10$ keV. Photons with lower energies are absorbed
locally by dust in the star forming region while more energetic
photons have such long mean-free-paths that they lose their energy to
redshift effects and some are not absorbed even by the end of
reionization. The photons that produce early cosmic heating might also
contribute to the unresolved soft X-ray background observed by {\it
  Chandra} \citep{Lehmer:2012} which can be used to put upper limits
on the efficiency of X-ray sources \citep{Fialkov:2016}.

As mentioned in the introduction, the most plausible sources for
dominating high redshift X-ray emission are XRBs. They are expected to
have a hard X-ray SED that peaks at about $1-3$ keV and is nearly
independent of redshift. We adopt the hard SED case from
\citet{Fragos:2013,Fialkov:2014b} to describe the spectral shape of
XRBs. Another category of possible X-ray sources which we consider
here are mini-quasars. Because their hard SED is similar in shape to
that of XRBs, with only a weak dependence on the black hole mass and
redshift \citep{Tanaka:2012}, we adopt the same shape of SED for
mini-quasars as for the XRBs for simplicity. In order to cover a wide
range of SEDs, we also consider the possibility of a soft power-law
spectrum, which is often used in the literature
\citep{Furlanetto:2006,Mesinger:2011}. The main difference between the
soft and the hard SEDs is that in the latter case the typical mean
free path of X-ray photons is much larger, so there is a long delay in
the energy absorption; the delay causes energy loss due to redshift
effects, so that the total absorbed energy is reduced by a factor of
$\sim5$ compared to the soft case \citep{Fialkov:2014b,Fialkov:2014c};
furthermore, hard sources at high redshifts tend to contribute more
photons within the energy range corresponding to the observed soft
X-ray background\citep{Fialkov:2016}.

To calculate the total X-ray luminosity we use the observed
SFR$-L_{X}$ relation:
\begin{equation}
\label{eq:XraySFR}
\frac{L_X}{\rm SFR}=3\times10^{40}f_X  \rm~erg~s^{-1}~M_\odot^{-1}~yr\ ,
\end{equation}
where $L_X$ is the bolometric luminosity summed over $0.2-95$ keV, and
$f_X$ is the X-ray efficiency of sources (assumed to be
constant). This relation is based on observations of nearby starburst
galaxies and XRBs \citep{Grimm:2003,Gilfanov:2004,Mineo:2012}, and the
standard normalization for XRBs (with $f_X=1$) includes an
order-of-magnitude increase in this ratio at the low metallicity
expected for high-redshift galaxies \citep{Fragos:2013}. In any case,
we try a wide range of values of $f_X$, so for us
Eq.~(\ref{eq:XraySFR}) is just a fiducial value. We use
Eq.~(\ref{eq:XraySFR}) for the cases of a hard or soft spectrum, and
in the case of mini-quasars we add to it the ratio between the X-ray
luminosity of XRBs (assumed given by Eq.~\ref{eq:XraySFR}) and that of
mini-quasars \citep{Wyithe:2003,Fialkov:2014b,Fialkov:2016}:
\begin{equation}
\frac{L_{\rm MQ}}{L_{\rm XRB}}\sim 0.1 \left(\frac{0.05}{f_*} \right)
\left(\frac{M_{\rm halo}}{10^8 M_\odot} \right)^{2/3}
\left(\frac{1+z}{10} \right)\ .
\end{equation}
The additional dependence of the mini-quasar luminosity on the halo
mass results in a relatively small contribution from these sources at
redshifts $z\gg8$, when halo masses are typically small, but they
become dominant (when mixed together with other sources) at lower
redshift, when larger halos form. Note that the luminosity of
mini-quasars includes the factor $f_X$ as well; in cases that include
two different X-ray sources the values of $f_X$ throughout this paper
indicate the total, where each population gets a normalization factor
equal to half the total.

Existing measurements can be used to constrain the value of $f_X$ for
each type of source.  \citet{Fialkov:2016} found that the unresolved
soft X-ray background gives an upper limit for $f_X$ that varies
between $\sim10-190$ depending on the nature of the X-ray sources, the
halo cooling channel and the value of the total CMB optical depth. A
lower limit on $f_X$ comes from measured upper limits on the 21-cm
power spectrum \citep{Ali:2015} (which will be discussed below);
\citet{Fialkov:2016} found limits in the range of $0-0.036$ (i.e.,
with some models unconstrained). In this paper, we take $f_X=1$ as our
standard value and mainly explore values in the range of $0.1-10$, but
consider also the extreme lower and upper limits from
\citet{Fialkov:2016}.

\subsection{CMB Optical Depth and Mean Free Path of Ionizing
    Photons}

The intensity of the 21-cm signal is proportional to the fraction of
neutral hydrogen atoms in the IGM (Eq.~\ref{eq:signal}), which is
determined by the progress of cosmic reionization. According to
current understanding, reionization happens inside out, proceeding
first in the dense regions containing most of the sources
\citep{Barkana:2004,Furlanettoetal:2004,Iliev:2006}. The amplitude of
the global 21-cm signal decreases as reionization advances.

A parameter that measures the total column density of ionized gas is
the total CMB optical depth, $\tau$. The latest and most precise
constraints on $\tau$ come from the {\it Planck} satellite
\citep{Planck:2015, Planck:2016a, Planck:2016b}. However, the error in
$\tau$ is still fairly large; moreover, the measured value of $\tau$
has gone down over time, and low values of $\tau$ are harder to
measure since their imprint on the CMB is weaker. In particular, data
analysis in 2015 (when we started our work) found an optical depth of
$\tau=0.066\pm0.016$ \citep{Planck:2015}, which was smaller than
previously measured; 2016 data gave $\tau=0.055\pm0.009$
\citep{Planck:2016b}, while in a companion paper \citep{Planck:2016a},
which used a more realistic reionization model, a slightly higher
value and uncertainty were reported, $\tau=0.058\pm0.012$. Keeping in
mind that in this work our main concern is to explore the widest
parameter space possible, we adopt $\tau = 0.066$ as our standard
value (it is difficult for us to produce much lower $\tau$ than this
and still complete reionization by $z \sim 6$), and also consider
$\tau = 0.082$ and $\tau = 0.098$, which are 1~$\sigma$ and 2~$\sigma$
away from the value reported by \citet{Planck:2015}. Because
$\tau>0.09$ is unlikely in the light of the latest results (i.e., it
is ruled out at about the 3-$\sigma$ level), in the next section we
mark all cases with $\tau>0.09$ differently in the figures and we
excluded these models from the fitting formulae.

The course of reionization depends also on the mean free path
  of ionizing photons. In particular, propagation of ionizing photons
  within the ionized regions is affected by the presence of absorption
  systems (Lyman-limit systems). We followed others
  \citep[e.g.,][]{Rmfp2} in modeling this effect by imposing an upper
  limit on the mean free path of ionizing photons inside ionized
  regions, $R_{\rm mfp}$.  We used the default value of $R_{\rm
    mfp}=70$ Mpc (comoving) for most of the cases [this represents the
    maximum value expected and perhaps observed near the end of
    reionization \citep{RmfpEnd}] but considered also the values of 20
  and 5 Mpc.  

\subsection{The Parameter Space}
\label{Sec:paramS}

In this section we define the space of the discussed astrophysical
parameters with the aim to (i) probe the range of possibilities for
the global 21-cm signal and (ii) reasonably fill up the space within
these boundaries. We stress that in this work we do not try to define
the probability of each parameter set because the high-redshift
universe is so poorly constrained. Before the 21-cm signal is
detected, it is important to stay open-minded and allow for any
reasonable realization within the allowed range.

We chose one model as a reference and refer to it as our ``standard''
case. For this case we used the atomic cooling scenario with
$f_*=0.05$, $f_X=1$, XRBs as the (hard) X-ray sources and
$\tau=0.066$. Next, we considered small and large variations in the
values of astrophysical parameters around the reference set by varying
each parameter and considering all possible combinations. For the
small variations (32 different parameter sets) we used either
molecular cooling or massive halos with $f_*$ and $f_X$ either larger
or smaller by a factor of $\sqrt{10}$ from their standard values, our
X-ray sources were either a mixture of XRBs and mini-quasars or a
mixture of a soft SED and mini-quasars, and we assumed either $\tau =
0.066$ or $\tau = 0.082$. For the large variations we took either
metal cooling or super-massive halos with $f_*$ and $f_X$ either
larger or smaller by a factor of $10$ from their standard values,
either X-ray sources with a soft SED or mini-quasars, and either
$\tau=0.066$ or $0.098$. This gave us 20 additional models (since
others were ruled out by the above observations or the inability to
achieve the desired $\tau$). In terms of the X-ray sources, we note
that mini-quasars give the case that is most different from a soft
SED, since mini-quasars not only have a hard X-ray spectrum (which
leads to weak early heating) but also decline faster with redshift
(which weakens early heating even more). Thus, the SED cases in the
``Small Variations'' category were roughly chosen to be intermediate
compared to the ``Large Variations'' and the ``Standard Case''. In
addition, in order to fill up the global signal space with more
intermediate models, we used all the combinations of the following
parameters: $\tau=[0.066,0.082]$, $f_*=[0.005,0.05,0.5]$,
$f_X=[0.1,1,8]$, $V_c=[4.2,16.5,35.5]$ km s$^{-1}$ and SED =
[soft,hard], which yielded 106 additional ``Space filler'' parameter
sets. Here the high $f_X$ value was chosen to be 8 rather than 10 in
order to allow us to normalize all the models to $\tau$ as low as
$6.6\%$. For all the cases described above we used $R_{\rm
    mfp}=70$ Mpc.  In addition, we considered some of the extreme
  cases with $R_{\rm mfp}=20$ and 5~Mpc, in order to widen the total
  parameter space.

To obtain the desired optical depth in each case, we set the requisite
value of the ionizing efficiency 
\citep{Barkana:2001,Furlanettoetal:2006}:
\begin{equation} \label{eq:zeta}
\zeta = f_*f_{\rm esc}N_{\rm ion}\frac{1}{1+\bar{n}_{\rm rec}}
\end{equation}
where $f_{\rm esc}$ is the fraction of ionizing photons that escape
into the IGM, $N_{\rm ion}$ is the mean number of ionizing photons
produced per stellar baryon and $\bar{n}_{\rm rec}$ is the mean number
of recombinations per ionized hydrogen atom. For each set of
parameters we tune $\zeta$ to produce the required optical depth while
requiring the ionization fraction to be at least 95\% at $z = 6$. An
upper limit on this parameter is $\zeta_{\rm max}=40,000 f_*$, where
we use the value of $N_{\rm ion}=40,000$ for massive Pop III stars
\citep{Bromm:2001} (We note that our models are based on numerical
values for Pop II stars, but by varying $f_*$ and $\zeta$ we
effectively cover a wide range of possibilities including the case of
massive Pop III stars). 

Some of our initial models were in conflict with recent upper limits
on the 21-cm power spectrum reported by the PAPER collaboration
\citep{Ali:2015}. These data rule out models in which the 21-cm
fluctuations at $z=8.4$ are larger than 22~mK in the range of
wavenumbers $k=0.15-0.5$ h Mpc$^{-1}$, where h is the Hubble constant
in units of 100 km s$^{-1}$ Mpc$^{-1}$. Note that using this result to
constrain the global signal directly is very model-dependent
(\citet{Pober:2015} applied it for a specific model). To make use of
this constraint we calculated the power spectrum for each case
separately. Applying the optical depth normalization and the
constraints from the PAPER experiment on our parameter space resulted
in some large variation cases that had to be modified or excluded,
leaving us with 20 such models.

In order to better explore the boundaries of the parameter space, we
also added 22 different cases presented in \citet{Fialkov:2016} (those
not ruled out by the PAPER upper limits). In that paper the optical
depth was not a free parameter but was instead determined by the
redshift at which reionization ends (either late reionization with
$z_{\rm re} =6.2$ or early reionization with $z_{\rm re} = 8.5$). The
scenarios of atomic and massive cooling with $f_*=0.05$ were
considered separately for each of the three types of X-ray sources
(XRBs, mini-quasars and soft power-law), while three different values
were assigned for $f_X$ --- the upper limit defined by the unresolved
soft X-ray background, the standard value of $f_X = 1$, and the lower
limit defined by the PAPER constraint which applies only in the case
of late reionization while for the early reionization scenario the
lower limit on X-ray efficiency was $f_X =0$.  A summary of the
parameter space is presented in Table \ref{table:variations}, and the
full list of cases is presented in Appendix~\ref{appA}.

\section{Results}
\label{Sec:results}

The parameter sets laid out in the previous section yield 193 total
cases of the global 21-cm signal shown in
Figure~\ref{fig:T21cmStacked}, where the black line corresponds to the
standard case with its three key turning points highlighted in
red. This figure demonstrates how the large uncertainty in the
astrophysical parameters introduced in Section~\ref{Sec:Param}
translates into a range of possibilities for the expected global 21-cm
signal. The signal is sensitive to each of the variable astrophysical
parameters and, thus, will have the power to constrain high-redshift
astrophysics once it is detected.

Figure~\ref{fig:T21cmStacked} yields the first important conclusion of
our parameter study: All the curves have the same basic qualitative
shape as our standard case. The quantitative positions (in $\nu$ and
$T_{\rm b}$) of the various key points vary among the curves, but the
overall structure is fixed. We can intuitively understand this as
follows. The low optical depth from Planck basically fixes
reionization to occur at the low end of the redshift range we cover,
with reionization completing at $z\sim 6-9$. At the other end, the
\Lya intensity required to produce \Lya coupling is rather low, so
that stars (assuming they contribute significantly to reionization)
naturally saturate \Lya coupling long before a significant fraction of
the Universe is reionized. As for X-ray heating, it can occur at a
wide range of redshifts, much earlier than reionization in the case of
strong heating with a soft SED and low optical depth, or as late as
the very end of reionization for scenarios with hard X-ray sources and
reionization at the high end of the optical depth range consistent
with Planck data \citep{Fialkov:2014b, Fialkov:2014c, Fialkov:2016b,
  Madau:2016, Mirocha:2016}. Importantly, though, X-ray heating always
begins somewhat {\em after}\/ the beginning of significant \Lya
coupling. Our results are thus reassuring, but they do not imply that
models that violate this basic shape are completely ruled out, though
such models do appear highly unlikely. For example, extremely strong
X-ray heating could occur prior to \Lya coupling (i.e., at $z \sim
30$) and prevent an absorption minimum, but in order for such an
intense X-ray outburst to avoid overproducing the observed X-ray
background, the associated source population would have to essentially
disappear by $z \sim 10$ \citep{Fialkov:2016} despite the rapid
ramp-up of galaxy formation. We also note that a different version of
Figure~\ref{fig:T21cmStacked} is shown later, at the end of this
section.

In this section we use our 193 different cases to explore the
correlation between the features of the global 21-cm signal and
physical properties of the high-redshift universe, showing that the
neutral hydrogen signal alone has enough predictive power to constrain
some details of primordial star formation, heating and reionization as
well as the typical halo mass. We analyze the properties of the 21-cm
signal starting from high redshifts at which the Universe was cold and
empty (and thus easy to analyze), then continuing to lower redshifts
at which various astrophysical processes caused non-local feedback
effects on star formation, complicating the picture.

\subsection{High-$z$ Maximum Point}

As noted in Section \ref{Sec:Methods}, the high-redshift maximum of
the global 21-cm curve is the point at which the first population of
Ly$\alpha$ sources (assumed to be stars) turns on, and the
Wouthuysen-Field coupling starts to become effective. At this point
the Universe is still relatively simple, as stars are rare and have
not yet had a significant affect on the 21-cm intensity. The only
parameters that have an effect on the 21-cm signal at this point are
the minimum mass of star forming halos and the star formation
efficiency, which together determine the Ly$\alpha$ intensity and thus
the strength of the Ly$\alpha$ coupling. The parameters related to
heating and reionization are not yet important. In the absence of
collisions and Ly$\alpha$ photons, interactions with the CMB would
drive $T_S$ to $T_{\rm CMB}$ and thus the differential brightness
temperature would be zero. At $z \sim 35$, collisions, which had kept
$T_s$ close to $T_{\rm gas}$ at higher redshift, become less and less
effective with time, so that $T_{\rm b}$ (which is negative) rises towards
zero. This portion of the global 21-cm curve is still within the
``dark ages'', i.e., precisely predictable given the basic
cosmological parameters.

\begin{figure}
	\centering
	\includegraphics[width=3in]{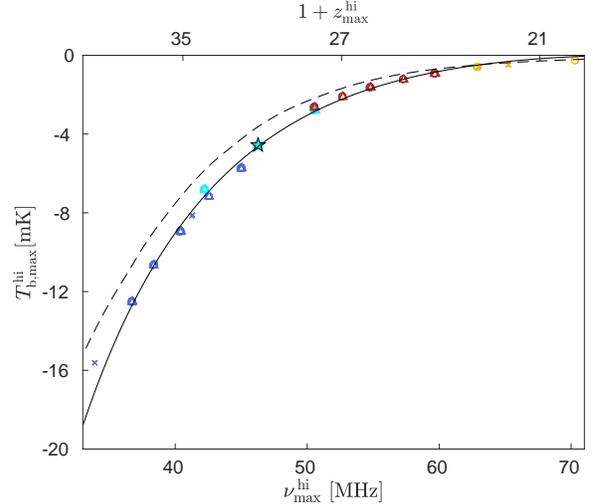}
	\caption{Brightness temperature as a function of observed
          frequency (bottom axis) or the equivalent one plus redshift
          (top axis) at the high-$z$ maximum point. The colors
          indicate the minimum circular velocity of star-forming halos
          for each case: $V_c=4.2$ (blue), 16.5 (cyan), 35.5 (red),
          and 76.5 km/s (yellow). Shapes indicate the optical
            depth for each case: $\tau=0.060 - 0.075$ (circles),
            $0.082 - 0.09$ (triangles), $0.09 - 0.111$ (crosses),
            while the star is our standard case. Also shown are the
          fitting function of Eq.~(\ref{eq:highz}) (solid curve), and
          the dark-ages (i.e., no astrophysical radiation) relation
          (dashed curve). }
	\label{fig:FirstMax21cm}
\end{figure}

The high-redshift maximum is produced just as the first significant
\Lya coupling causes $T_{\rm b}$ to start becoming more negative. Thus, the
high-$z$ maximum occurs near the dark ages curve, so we expect to see
an approximate relation between the redshift of the maximal point,
$z_{\rm max}^{\rm hi}$, and the value of the brightness temperature at
this point, $T_{\rm b,max}^{\rm hi}$; specifically the lower the
$z_{\rm max}^{\rm hi}$, the closer $T_{\rm b,max}^{\rm hi}$ should be
to zero. We indeed observe a rather clear relation in our models, as
shown in Figure~\ref{fig:FirstMax21cm}. We find the following
approximate fitting formula:
\begin{equation}
\label{eq:highz}
T_{\rm b,max}^{\rm hi}=a(1+z_{\rm max}^{\rm hi})^2+b(1+z_{\rm
  max}^{\rm hi})+c\ ,
\end{equation}
where $[a,b,c]=[-0.03124,1.155,-10.65]$ (We note that
\citet{Mirocha:2013} used an analytical analysis for an approximate
study of this dependence). At the highest redshifts, the dark-ages
global 21-cm curve rises steeply, so that it takes some time for it to
turn over due to \Lya radiation, and the maximum point occurs below
the dark-ages curve by $\sim 2$~mK at a given redshift; at lower
redshifts, the dark-ages curve is flatter so that the maximum occurs
almost immediately once the actual curve deviates from the dark-ages
limit.

Since the relation between $T_{\rm b,max}^{\rm hi}$ and $z_{\rm
  max}^{\rm hi}$ is monotonic and there is almost no scatter, it would
suffice to measure either the brightness temperature or the redshift
to obtain all the information on this extremum of the global signal
(though measuring both would provide a clear consistency test and
verification that the expected signal is indeed being observed). From
this measurement it should be possible to roughly estimate the value
of the minimum $V_c$ (shown by different colors in the Figure), with
an uncertainty introduced by the possible range of the SFE. As
expected, smaller $V_c$ implies earlier the star formation and thus a
more negative value of $T_{\rm b,max}^{\rm hi}$.

\begin{figure}
	\centering
	\includegraphics[width=3in]{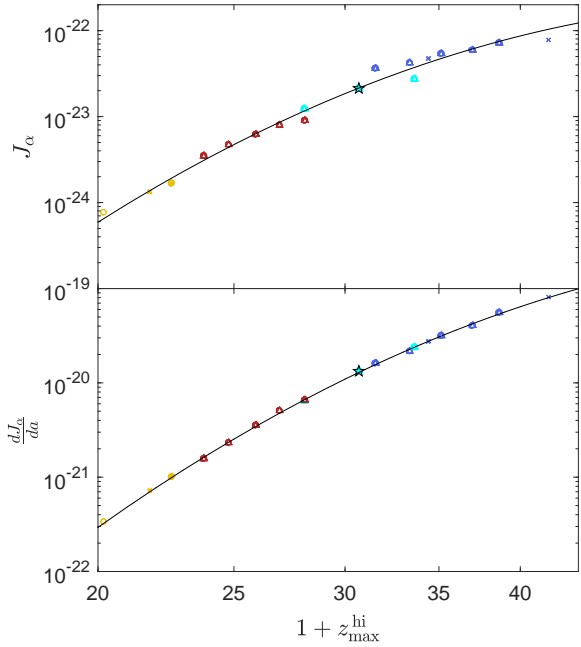}
	\caption{The Ly$\alpha$ intensity (top panel) in units of erg
          s$^{-1}$ cm$^{-2}$ Hz$^{-1}$ sr$^{-1}$ and its derivative
          with respect to the scale factor (bottom panel) as a
          function of $z_{\rm max}^{\rm hi}$. The color indicates the
          cooling channel for each case: $V_c=4.2$ (blue), 16.5
          (cyan), 35.5 (red), 76.5 km s$^{-1}$ (yellow). Shapes indicate the optical depth for each case: $\tau=0.060 -
          	0.075$ (circles), $0.082 - 0.09$ (triangles), $0.09 - 0.111$
          	(crosses), while the star is our standard case. Also shown (black lines) are the fitting
          function Eq.~(\ref{eq:JA}) (top panel), and
          Eq.~(\ref{eq:dJA}) (bottom panel).}
	\label{fig:JA}
\end{figure}

In addition to considering predicted relations among observables of
the global 21-cm spectrum, our other goal in this paper is to explore
whether astrophysical information can be easily extracted. Additional
information that could be extracted the high-$z$ maximum were measured
is the average intensity of the Ly$\alpha$ background at this epoch,
i.e., at redshift $z_{\rm max}^{\rm hi}$. As can be seen from
Figure~\ref{fig:JA}, the observables can be used to accurately
reconstruct the angle-averaged intensity of Ly$\alpha$ photons,
$J_\alpha$, as well as its derivative with respect to the scale
factor, $a$ (both spatially-averaged over the universe). This is true
whether we use the redshift, as shown, or the brightness temperature
(which is strongly correlated with it based on
Figure~\ref{fig:FirstMax21cm})). 

We fit the dependence with the following functions (hereafter, $\log$
means base 10):
\begin{equation}
\label{eq:JA}
\log\left( J_\alpha\right) =a_1\log^2\left( 1+z_{\rm max}^{\rm
  hi}\right) +b_1\log\left( 1+z_{\rm max}^{\rm hi}\right) +c_1\ ,
\end{equation}
\begin{equation}
\label{eq:dJA}
\log\left( \frac{dJ_\alpha}{da}\right) =a_2\log^2\left( 1+z_{\rm
  max}^{\rm hi}\right) +b_2\log\left( 1+z_{\rm max}^{\rm hi}\right)
+c_2\ ,
\end{equation}
where $[a_1,b_1,c_1]=[-10.39,37.36,-55.247]$ and
$[a_2,b_2,c_2]=[-9.238,34.59,-50.897]$, and $J_\alpha$ is in units of
erg s$^{-1}$ cm$^{-2}$ Hz$^{-1}$ sr$^{-1}$. Figure~\ref{fig:JA} shows
that the scatter in these relations is fairly small, especially for
the derivative of the intensity, since the rapid change in time in
$J_\alpha$ makes the extremum condition especially sensitive to the
derivative. If the time derivative of the Ly$\alpha$ intensity is
determined in this way, this would provide information on a
combination of the minimum halo $V_c$ and the star formation
efficiency.

\subsection{Minimum Point}

\begin{figure}
\centering
\includegraphics[width=3in]{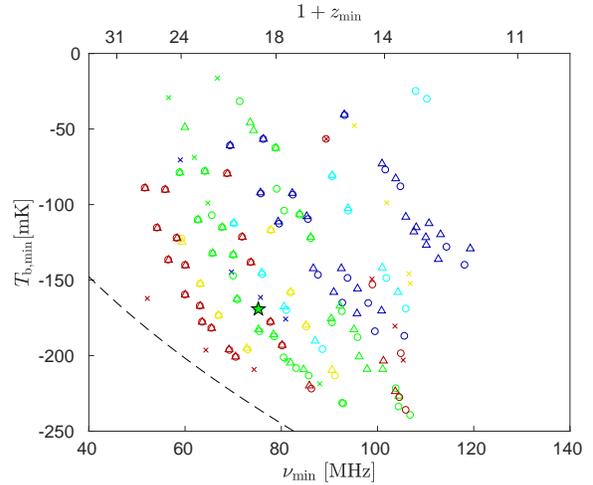}
\caption{Brightness temperature at the minimum point as a function of
  observed frequency of this point (bottom axis) or the equivalent one
  plus redshift (top axis). The colors indicate the star formation
  efficiency for each case: $f_*=0.005$ (blue), 0.016 (cyan), 0.05 (green), 0.16
  (yellow), 0.5 (red). Also shown (black dashed line) is the lower limit on the
  brightness temperature at each redshift from the no-heating limit
  (Eq.~\ref{eq:minLim}). Shapes indicate the optical depth for each case: $\tau=0.060 -
  	0.075$ (circles), $0.082 - 0.09$ (triangles), $0.09 - 0.111$
  	(crosses), while the star is our standard case. }
\label{fig:Min21cm}
\end{figure}

The next prominent feature of the global 21-cm signal is the
absorption trough, with the minimum defined by the beginning of the
heating era in combination with \Lya saturation. The location of this
point depends on more parameters (the X-ray efficiency and SED in
addition to $V_c$ and $f_*$), which leads to a larger scatter of
possible values of the minimal brightness temperature $T_{\rm b,min}$
and the redshift at which it is achieved $z_{\rm min}$. While for our
standard case the minimum point occurs at $z_{\rm min}=18$ with a
depth of $T_{\rm b,min}=-170$ mK, variation of the astrophysical
parameters leads to a range in redshift $10.9<z_{\rm min}<26.5$ and
temperatures $-240$ mK $<T_{\rm b,min}<-25$ mK. As can be seen in
Figure~\ref{fig:Min21cm}, there is no predicted relation between these
two observables, the redshift and the temperature, e.g., for any given
redshift of the minimum a large range of corresponding temperatures is
possible.

We can understand how some of the parameters produce the large scatter
in the plot. Points with the same $f_*$ (marked by the same color in
Figure~\ref{fig:Min21cm}) are roughly aligned along diagonal lines
going from the top left to the bottom right of the plot. Different
lines of the same color correspond to different values of $V_c$, while
the scattering of points along a given line is due to variations in
the intensity and SED of the X-ray radiation. Lower values of $V_c$
lead to higher $z_{\rm min}$, while a lower X-ray heating rate results
in a more negative value of $T_{\rm b,min}$.

A lower limit on the brightness temperature can be obtained assuming a
fully neutral universe with no X-ray sources (i.e., where the gas
cools adiabatically after thermal decoupling from the CMB) but full
\Lya coupling (i.e., $T_S = T_{\rm gas}$). In this limit, using
Eq.~(\ref{eq:signal}) we can derive the following relation
for the mean temperature:
\begin{equation}
\label{eq:minLim}
T_{\rm b,min}\geq26.8\left(\frac{1+z_{\rm min}}{10} \right)^{1/2}
\left( 1- \frac{1+z_{\rm dec}}{1+z_{\rm min}}\right) \rm mK
\end{equation}
where $z_{\rm dec}=137$ is the redshift at which the gas temperature
and the CMB temperature effectively decoupled. Following the same
logic, we can write
\begin{equation}
\label{eq:TcmbDTgas}
1-\frac{T_{\rm b,min}}{26.8\sqrt{\frac{1+z_{\rm min}}{10}}} \leq
\frac{T_{\rm CMB}}{T_{\rm gas}} \leq \frac{1+z_{\rm dec}}{1+z_{\rm
    min}}\ .
\end{equation}

\begin{figure*}
	\centering
	\includegraphics[width=6in]{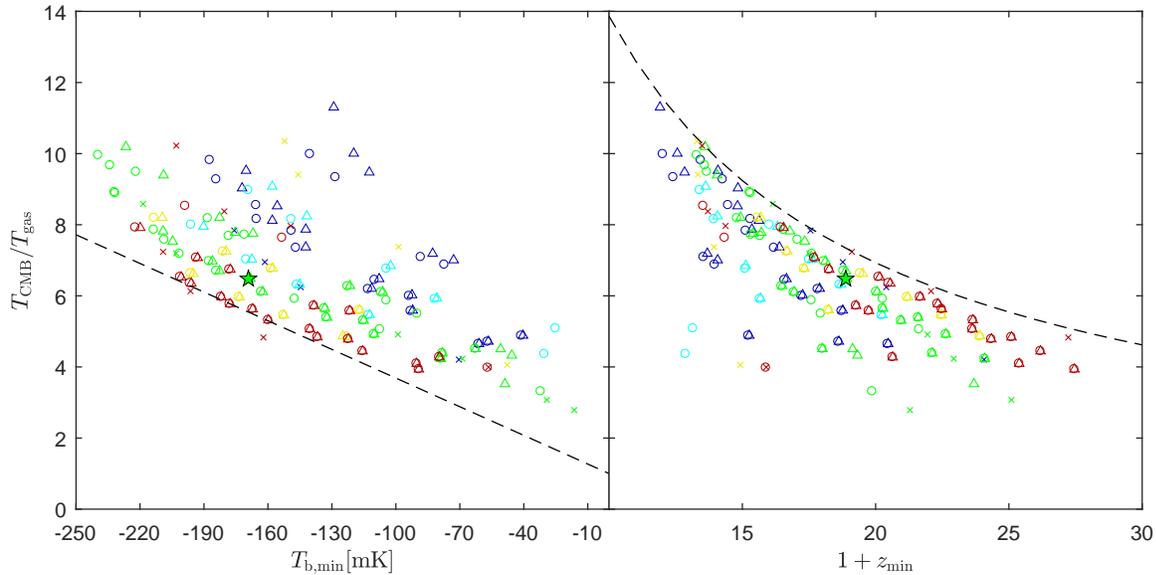}
	\caption{The ratio $T_{\rm CMB}/T_{\rm gas}$ as a function of
          the brightness temperature (left panel) and of the redshift
          (right panel) at the minimum point. The colors indicate the
          star formation efficiency for each case: $f_*=0.005$ (blue),
          0.016 (cyan), 0.05 (green), 0.16 (yellow), 0.5 (red). Shapes indicate the optical depth for each case: $\tau=0.060 -
          	0.075$ (circles), $0.082 - 0.09$ (triangles), $0.09 - 0.111$
          	(crosses), while the star is our standard case. Also
          shown (dashed black lines) is the lower limit on $T_{\rm
            CMB}/T_{\rm gas}$ (left panel) and upper limit on $T_{\rm
            CMB}/T_{\rm gas}$ (right panel); see
          Eq.~(\ref{eq:TcmbDTgas}) and the text for details.}
	\label{fig:Tcmb}
\end{figure*}

Figure~\ref{fig:Tcmb} shows the ratio $T_{\rm CMB}/T_{\rm gas}$ as a
function of the brightness temperature or the redshift at the minimum
point. The black dashed line in the left panel shows the left-hand
side of Eq.~(\ref{eq:TcmbDTgas}) calculated at $z_{\rm min}=18$ which
corresponds to $z_{\rm min}$ in our standard case. Since the redshift
dependence is weak, taking a constant redshift is a good approximation
to the case where the spin temperature is fully coupled to the gas
temperature (note also that the reionized fraction is quite low at
this stage in our models). While many of our simulated models lie
close to this line, implying that they have nearly achieved saturated
\Lya coupling, a substantial fraction are well away from the line,
showing that for them the Wouthuysen-Field coupling is still far from
saturation. The strength of the coupling is determined by the
Ly$\alpha$ intensity and is stronger for models with large $f_*$ and
small $V_c$. The black dashed line in the right panel of
Figure~\ref{fig:Tcmb} shows the right-hand side of
Eq.~(\ref{eq:TcmbDTgas}). Here too, while many models show little
heating (i.e., are close to the line), a substantial fraction have
undergone significant heating prior to reaching the minimum. Even when
heating begins, in order to produce a global 21-cm minimum it must
overcome two effects: adiabatic cooling (astrophysical heating
initially only slows down the rate of cooling), and the increase with
time of the \Lya coupling (which, as long as it is not yet saturated,
pushes $T_{\rm b}$ to be more negative). The two panels of
Figure~\ref{fig:Tcmb} together show the large variety of physical
conditions that leads to the large scatter in the position of the
minimum point.

\begin{figure}
	\centering
	\includegraphics[width=3in]{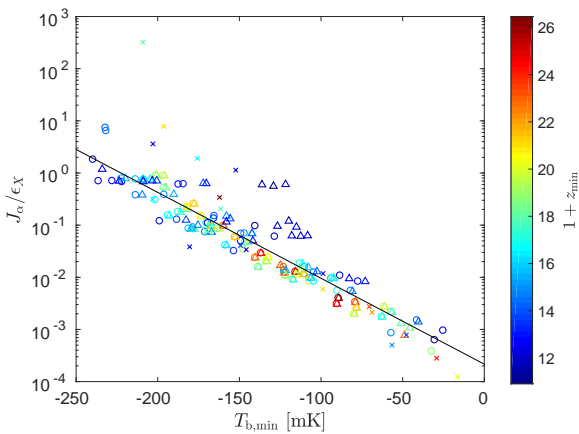}
	\caption{The ratio between the Ly$\alpha$ intensity (in units
          of erg s$^{-1}$ cm$^{-2}$ Hz$^{-1}$ sr$^{-1}$) and the X-ray
          heating rate (in units of eV s$^{-1}$ baryon$^{-1}$) as a
          function of the brightness temperature at the minimum
          point. The color indicates the redshift of the minimum point
          (see the color bar on the right). Also shown is the fitting
          function of Eq.~(\ref{eq:minJdEPS}) (black line). Shapes indicate the optical depth for each case: $\tau=0.060 -
          	0.075$ (circles), $0.082 - 0.09$ (triangles), $0.09 - 0.111$
          	(crosses), while the star is our standard case. Note that the point with the largest ratio
          represent an extreme case in which X-ray sources are
          mini-quasars and $f_X=0.1$, which means an extremely low
          X-ray heating rate at this redshift, while metal cooling
          with $f_*=0.5$ drives up $J_\alpha$.  }
	\label{fig:minT21cmJalphaDepsXT}
\end{figure}

Despite the complexity, we have managed to find some order in the
relation between the properties of the minimum and the underlying
astrophysical properties of interest. Specifically, the depth of the
minimum point is strongly correlated with the ratio between the
average Ly$\alpha$ intensity and the X-ray heating rate, as shown in
Figure~\ref{fig:minT21cmJalphaDepsXT}. The correlation is easy to
explain: with a large X-ray heating rate, heating starts earlier and
the brightness temperature at the minimum (the redshift of which
depends also on \Lya saturation) is less negative. On the other hand,
large Ly$\alpha$ intensity leads to stronger coupling between the spin
temperature and the gas temperature, and, thus, to a more negative
brightness temperature when significant heating begins and produces
the minimum. We also show in Figure~\ref{fig:minT21cmJalphaDepsXT} a
fitting function (always excluding $\tau>0.09$ cases, as noted
previously):
\begin{equation}
\label{eq:minJdEPS}
\log\left( \frac{J_\alpha}{\epsilon_X}\right)  =aT_{\rm b,min}+b\ ,
\end{equation}
with $[a,b]=[-0.016,-3.666]$, where $\epsilon_X$ is the heating rate
in units of eV s$^{-1}$ per baryon in the IGM.  Thus, measuring the
minimum point, which is expected to be the most prominent feature of
the global 21-cm signal, will provide us a fairly good estimate of
$J_\alpha / \epsilon_X$ at the corresponding redshift.

\subsection{Low-$z$ Maximum Point}

The lowest-redshift extremum in the global 21-cm curve is the low-$z$
maximum. In models where X-ray sources are efficient and heat the gas
well above the temperature of the CMB early enough, the 21-cm signal
is seen in emission during the later stages of cosmic evolution.
Heating increases the emission signal until heating saturates ($T_{\rm
  gas}\gg T_{\rm CMB}$), producing another maximum in the global 21-cm
signal, of height $T_{\rm b,max}^{\rm lo}$ and redshift $z_{\rm
  max}^{\rm lo}$. The position of the maximum is also affected by the
advance of reionization. As more and more stars appear and galaxies
grow, the bubbles of reionized gas expand. Reionization suppresses the
neutral fraction and thus the intensity of the signal, helping produce
the maximum earlier while decreasing its brightness temperature. In
some models, including the currently most likely heating cases
\citep{Fialkov:2014b, Fialkov:2016b, Fialkov:2016, Mirocha:2016,
  Madau:2016}, heating is not saturated by the beginning of
reionization, the 21-cm signal is driven by heating and reionization
simultaneously, and the emission feature is not so
prominent. Furthermore, in cases of extremely low heating the gas in
neutral regions is colder than the CMB even at the end of reionization
\citep{Fialkov:2016b, Fialkov:2016} and should be seen in absorption
against the CMB at all redshifts, with no emission peak. In these
cases, where there is no low-$z$ maximum, we set $T_{\rm b,max}^{\rm
  lo}$ to zero and $z_{\rm max}^{\rm lo}$ to the end of reionization.

\begin{figure*}
	\centering \includegraphics[width=3in]{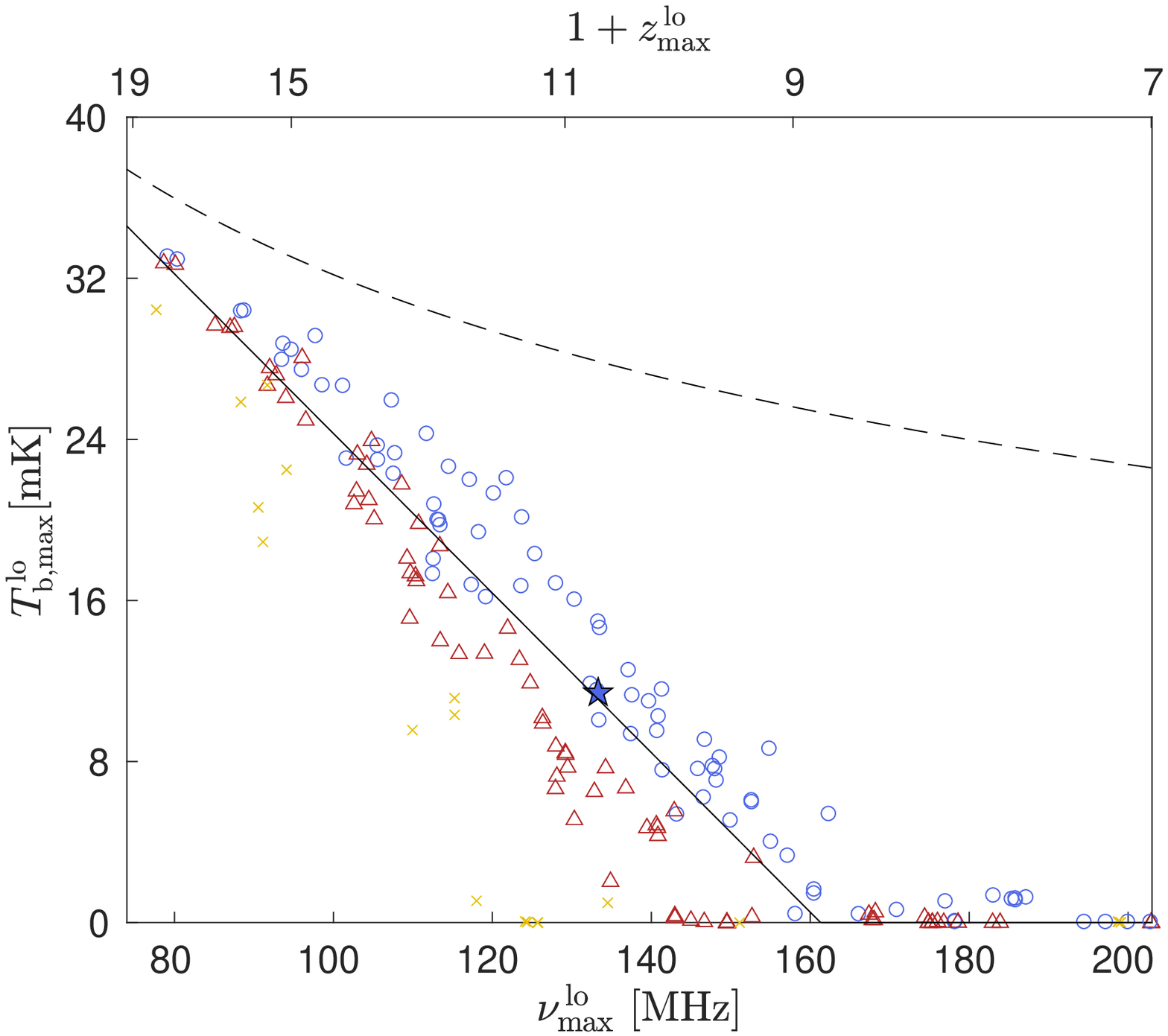}
	\hspace{0.4cm}
	\includegraphics[width=3.1in]{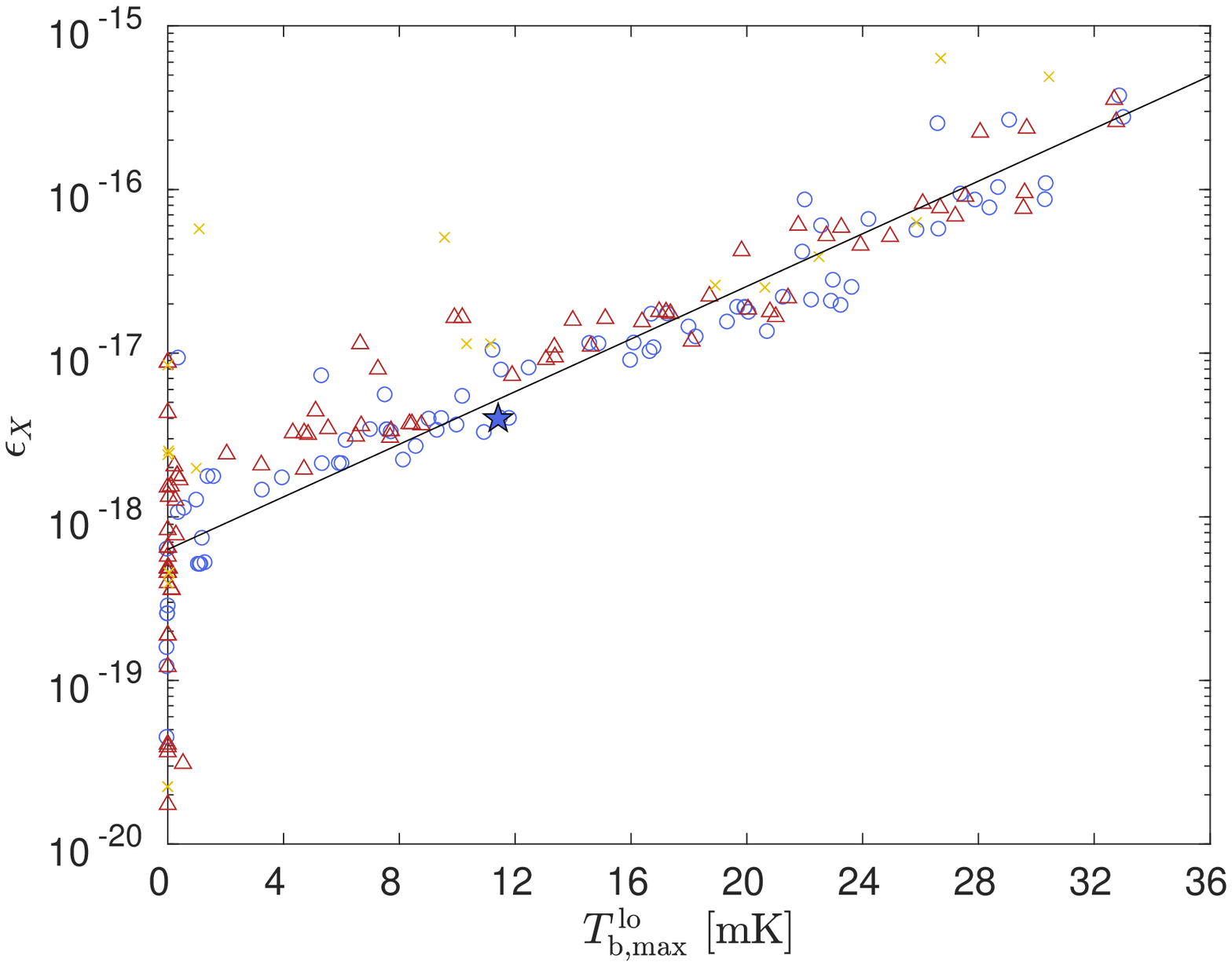}
	\caption{Left panel: Brightness temperature as a function of
          observed frequency (bottom axis) or equivalent one plus
          redshift (top axis) at the low-$z$ maximum point. Also shown
          is the fitting function given by Eq.~(\ref{eq:lowzTZ})
          (black solid line) and the upper limit from
          Eq.~(\ref{eq:maxLim}) (black dashed line). Right panel: The
          X-ray heating rate (in units of eV s$^{-1}$ baryon$^{-1}$)
          as a function of the brightness temperature at the low-$z$
          maximum. Also shown is the fitting function from
          Eq.~(\ref{eq:lowzZ}) (black line). In both panels the colors
          indicate the optical depth for each case: $\tau=0.060 -
          0.075$ (blue), $0.082 - 0.09$ (brown), $0.09 - 0.111$
          (yellow crosses); the star is our standard case.}
	\label{fig:maxlT-X}
\end{figure*}

The left panel of Figure~\ref{fig:maxlT-X} shows the scatter of the
low-redshift maximum point for all the considered models in the
$T_{\rm b,max}^{\rm lo}$ versus $\nu$ (or $z_{\rm max}^{\rm lo}$)
plane.  The emission signal cannot be stronger than the upper limit
obtained for a fully neutral universe ($x_{\rm HI}=1$) and saturated
heating:
\begin{equation}
\label{eq:maxLim}
T_{\rm b,max}^{\rm lo}=26.8\left(\frac{1+z_{\rm max}^{\rm lo}}{10}
\right)^{1/2} \rm mK.
\end{equation}
Usually heating is near saturation at the low-$z$ max, and the
distance between the model points in Figure~\ref{fig:maxlT-X} and the
dashed line (Eq.~(\ref{eq:maxLim})) mostly expresses the advance of
reionization towards low redshift. Current CMB observations restrict
our models to a fairly narrow range of reionization histories,
particularly if we restrict the optical depth to values consistent
with current measurements to within 2~$\sigma$ (roughly the blue
points in the figure). Moreover, the dependence on optical depth is
easy to explain: in cases with lower optical depth reionization starts
later and thus, at a given $z_{\rm max}^{\rm lo}$, it is less well
advanced, and the emission signal $T_{\rm b,max}^{\rm lo}$ is higher.

In general, models with strong heating tend to produce an emission
signal, and then saturated heating, at higher redshifts, already at an
early stage of reionization. In these cases the neutral fraction is
still high, and the value of $T_{\rm b,max}^{\rm lo}$ is closer to the
upper limit defined by Eq.~(\ref{eq:maxLim}). Extreme models with the
strongest X-ray emission feature a significant X-ray contribution to
reionization, which helps keep the high-redshift data points away from
the dashed line in Figure~\ref{fig:maxlT-X}. On the other hand, models
with weak heating need more time to arrive at the saturation point,
and thus the peak of the emission signal happens during advanced
stages of reionization by stellar sources ($x_{\rm HI}\ll 1$). Thus,
the amplitude of the emission maximum is much lower than the upper
bound and $z_{\rm max}^{\rm lo}$ is then closer to the end of
reionization. When considering the full ensemble of models, a roughly
linear dependence between the temperature and the frequency can be
seen (excluding the $T_{\rm b,max}^{\rm lo}=0$ points; left panel of
Figure~\ref{fig:maxlT-X}), which is well fitted by
\begin{equation}
\label{eq:lowzTZ}
T_{\rm b,max}^{\rm lo}=
\begin{cases}
a\frac{1}{1+z_{\rm max}^{\rm lo}}+b\ ,& \text{if } 1+z_{\rm max}^{\rm
  lo}>\frac{-a}{b}\ ,\\ 0\ , & \text{otherwise,}
\end{cases}
\end{equation}
where $[a,b]=[-562.8,63.9]$.

\begin{figure}
	\centering
	\includegraphics[width=3in]{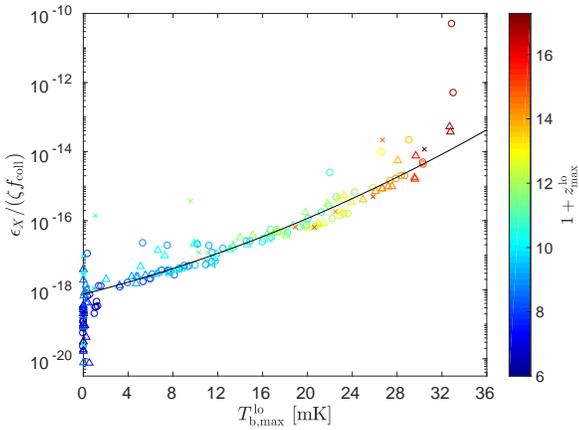}
	\caption{The ratio between the heating rate of X-ray sources,
          $\epsilon_X$ (in units of eV s$^{-1}$ baryon$^{-1}$), and
          the production of ionizing photons as measured by $\zeta
          f_{\rm coll}$, as a function of the brightness temperature
          at the low-$z$ maximum point. The color (see the color bar
          on the right) indicates the corresponding redshift of the
          low-$z$ maximum point. Also shown is the fitting function of
          Eq.~(\ref{eq:lowzZratio}) (solid black line). Shapes indicate the optical depth for each case: $\tau=0.060 -
          	0.075$ (circles), $0.082 - 0.09$ (triangles), $0.09 - 0.111$
          	(crosses), while the star is our standard case.}
	\label{fig:maxlT-epsDzetafcoll}
\end{figure}

We have found some interesting trends related to the low-redshift
emission point. Specifically, the strength of the emission signal can
be related to some of the astrophysical parameters at that epoch, thus
directly constraining heating sources and star formation. We find that
the intensity of the emission signal is correlated with the heating
rate at $z_{\rm max}^{\rm lo}$ (Figure~\ref{fig:maxlT-X} right panel),
with the relation well-fitted by:
\begin{equation}
\label{eq:lowzZ}
\log\left( \epsilon_X\right) =aT_{\rm b,max}^{\rm lo}+b\ ,
\end{equation}
where $[a,b]=[0.080,-18.2]$. The qualitative dependence is easy to
explain following the same lines as above: the larger the heating
rate, the earlier X-ray heating saturates and the stronger is the
emission feature in the absence of much reionization by UV
sources. Another striking correlation involves also the production of
ionizing photons. We use $\zeta f_{\rm coll}$ as a measure for the
produced amount of ionizing photons, where $f_{\rm coll}$ is the
fraction of mass in star forming halos (often called the ``collapsed
fraction''), which depends on $V_c$, and $\zeta$ is the overall
ionizing efficiency (Eq.~\ref{eq:zeta}). We plot the ratio of heating
rate to ionization production ($\epsilon_X/[\zeta f_{\rm coll}]$) as a
function of peak brightness temperature $T_{\rm b,max}^{\rm lo}$
(Figure~\ref{fig:maxlT-epsDzetafcoll}). Along the same lines of
reasoning as above, it is clear that to get large values of $T_{\rm
  max}^{\rm lo}$ a strong heating (which gives an early peak) together
with weak ionization (which keeps the neutral hydrogen fraction high)
is required. This relation can be fitted by:
\begin{equation}
\label{eq:lowzZratio}
\log\left( \frac{\epsilon_X}{\zeta f_{\rm coll}}\right) =a \left[
  T_{\rm b,max}^{\rm lo} \right]^2+b T_{\rm b,max}^{\rm lo}+c\ ,
\end{equation}
where $[a,b]=[0.0014,0.082,-18.13]$, and we excluded from the fit
points with $T_{\rm b,max}^{\rm lo}=0$ or $T_{\rm b,max}^{\rm
  lo}>32$~mK (or, as always, $\tau>.09$). Because the relation between
$T_{\rm b,max}^{\rm lo}$ and $z_{\rm max}^{\rm lo}$ is largely
monotonic as follows from Eq.~(\ref{eq:lowzTZ}) and is also indicated
by the color map in Figure~\ref{fig:maxlT-epsDzetafcoll}, it is also
possible to express the ratio as a function of $z_{\rm max}^{\rm lo}$
instead of the peak brightness temperature. Eqs.~(\ref{eq:lowzZ}) and
(\ref{eq:lowzZratio}) should enable us to estimate the X-ray heating
rate and $\zeta f_{\rm coll}$ at $z_{\rm max}^{\rm lo}$ from future
measurements of the global 21-cm signal.

\subsection{Average slopes}

\begin{figure}
	\centering
	\includegraphics[width=3.2in]{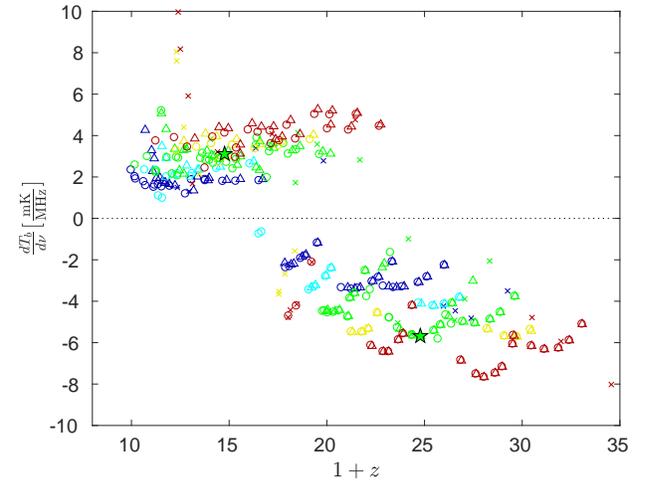}
	\caption{The mean slope of $T_{\rm b}$ versus $\nu$ between
          $z_{\rm max}^{\rm hi}$ and $z_{\rm min}$ (the negative
          slope) as a function of $1+\left(z_{\rm min}+z_{\rm
            max}^{\rm hi}\right)/2$, and the mean slope between
          $z_{\rm min}$ and $z_{\rm max}^{\rm lo}$ (the positive
          slope) as a function of $1+\left(z_{\rm min}+z_{\rm
            max}^{\rm lo}\right)/2$. Colors indicate different values
          of the star formation efficiency: $f_*$ = 0.005 (blue), 0.016
          (cyan), 0.05 (green), 0.16 (yellow), 0.5 (red). Shapes indicate the optical depth for each case: $\tau=0.060 -
          	0.075$ (circles), $0.082 - 0.09$ (triangles), $0.09 - 0.111$
          	(crosses), while the star is our standard case.}
	\label{fig:Slope}
\end{figure}

In addition to the three key points discussed above, the derivative
(slope) of the signal with respect to frequency is interesting to
consider separately, because it is likely easier to measure than the
absolute signal itself due to the need for foreground removal. We use
the key inflection points to define two characteristic slopes, for
each model: the mean slope between the high-$z$ maximum point and the
minimum point (a negative slope), and the mean slope between the
minimum point and the low-$z$ maximum point (a positive slope). This
allows us to visualize a kind of summary of the entire relevant range
of the global 21-cm curve, by plotting both the positive and negative
slopes together in Figure~\ref{fig:Slope}. Each slope is shown as a
function of the mean one plus redshift at which it is measured, i.e.,
$1+\left(z_{\rm min}+z_{\rm max}^{\rm hi}\right)/2$ for the negative
slope and $1+\left(z_{\rm min}+z_{\rm max}^{\rm lo}\right)/2$ for the
positive slope. 

Because each slope depends on the intensity and redshift of the global
signal at two of the turning points, the dependence on our various
parameters is more complex, and it is difficult to extract simple
relations between the slopes and astrophysical quantities. One general
trend is that a high star formation efficiency tends to produce strong
radiation fields early on, thus high redshifts for the key points
which imply shorter frequency intervals, resulting in steep slopes
(both positive and negative). The overall range of slopes is roughly
-1 to -8~mK/MHz (negative) and 1 to 5~mK/MHz (positive), with the
negative slope typically nearly twice the positive one (in absolute
value). Of course, the foreground emission is also substantially
larger at the higher redshift range corresponding to the negative
slope. Note that a very large positive slope ($dT_{\rm b}/d\nu \simgt
6$~mK/MHz) can only be produced by having reionization so early that
the resulting optical depth is excluded by Planck at 3~$\sigma$.

\subsection{Summary plot}

\begin{figure}
	\centering
	\includegraphics[width=3.2in]{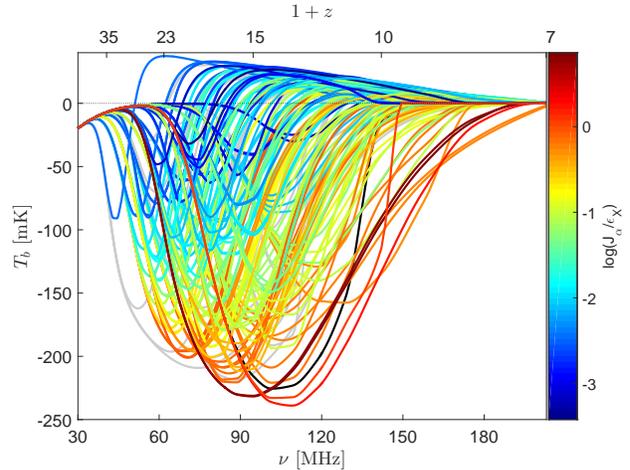}
	\caption{The 21-cm global signal as a function of redshift,
          for our 193 different astrophysical models, as in
          Figure~\ref{fig:T21cmStacked}. The color (see the color bar
          on the right) indicates the ratio between the Ly$\alpha$
          intensity (in units of erg s$^{-1}$ cm$^{-2}$ Hz$^{-1}$
          sr$^{-1}$) and the X-ray heating rate (in units of eV
          s$^{-1}$ baryon$^{-1}$) at the minimum point. Grey curves
          indicate cases with $\tau>0.09$, and a non-excluded $f_X=0$
          case is in black; these cases are all excluded from the
          color bar range.}
	\label{fig:Sum}
\end{figure}

Given the various results shown thus far in this section, in
particular the correlations between observable features and various
astrophysical parameters, we can construct a plot that partly
summarizes the correlations while showing the full global 21-cm
curves. Figure~\ref{fig:Sum} is another version of
Figure~\ref{fig:T21cmStacked}, but with additional information that
brings some order. First, cases with $\tau>0.09$ are shown as
grey. The remaining curves are color coded according to the
corresponding ratio in each model between the Ly$\alpha$ intensity and
the X-ray heating rate at $z_{\rm min}$ (i.e., at the time of the
minimum point of the global 21-cm curve). This is the same ratio
plotted in Figure~\ref{fig:minT21cmJalphaDepsXT}) and shown there to
correlate closely with the depth of the minimum, $T_{\rm
  b,min}$. While no single parameter can fully describe the global
21-cm curve, Figure~\ref{fig:Sum} does show that this particular ratio
nicely slices up the parameter space of possible curves. Those with a
high ratio tend to fill the bottom right portion of the figure, i.e.,
they usually produce a deep minimum and maintain a 21-cm absorption
signal until fairly late; models with a low ratio, on the other hand,
tend to fill the upper left, producing a shallow minimum and early
21-cm emission. Note that the color coding also makes it easier to
trace individual global 21-cm curves within this plot, in order to
appreciate the variety of curves that are possible. 

\section{Summary and Discussion}
\label{Sec:sum}

In this paper we have explored the allowed parameter space of the
global 21-cm signal, varying the main high-redshift astrophysical
parameters such as the minimal mass of star-forming halos, star
formation efficiency, and heating and ionization rates, all of which
are poorly constrained. The large uncertainty in high-redshift
astrophysical processes results in weak limits on the predicted 21-cm
signal. We used a realistic semi-numerical simulation to produce the
21-cm global signal in the redshift range $z=6-40$ for 193 different
sets of astrophysical parameters in agreement with current
observations (except that 21 are excluded by the Planck measurement of
optical depth at 3~$\sigma$). We applied these data to establish
universal patterns in the predicted global 21-cm curves. We found that
the general shape of the signal can be predicted theoretically, but
its features remain highly unconstrained. Still, there are clear
correlations between the three key features of the global 21-cm signal
(the high-$z$ maximum, the minimum and the low-$z$ maximum points) and
underlying astrophysical parameters of the early universe. Our
compilation of realistic models and fitting formulae for these
correlations can be used to rule out portions of the parameter space
as data from ongoing and future radio experiments becomes
available. If and when the global signal is measured, our results can
be used to reconstructed key aspects of the high-redshift population
of sources including the first stars, X-ray binaries, and
mini-quasars.

The parameters that we varied can be divided into three categories. The
first group consists of parameters related to primordial star
formation, including the minimum mass of halos in which stars can form
and the star formation efficiency. These parameters are the only ones
that affect the shape of the global signal (through the \Lya
intensity) from the formation of the first stars down to the redshift
where X-ray sources turn on. The second group captures properties of
the first heating sources, including their X-ray spectra, luminosity,
and evolution with redshift (e.g., XRBs versus mini-quasars), which
together with the properties of star formation affect the shape of the
signal from the moment when X-ray sources turn on to the point when
reionization becomes significant. Finally, the CMB optical depth is
related to ionization properties of stars which drive the global
signal at the low-redshift end.

As anticipated, properties of the high-$z$ maximum point are
relatively simple because it occurs at high redshift where significant
\Lya coupling begins, which in all our models occurs before heating
and ionizing sources complicate the evolution of the global
signal. There is a close relation between the redshift of this turning
point and the corresponding intensity of the 21-cm signal, providing a
potential consistency check for observations. These observable
quantities also correlate closely with the Ly$\alpha$ intensity and
its derivative at that epoch, according to fitting formulae that we
have obtained. Thus, measuring the global signal at this point would
help determine the total star formation rate at this early epoch, thus
constraining a combination of the minimum cooling mass of star forming
halos and the star formation efficiency.

The redshift and depth of the absorption trough show the largest
scatter, since this minimum can occur under various physical
conditions and is affected by many astrophysical parameters. It
typically occurs when \Lya coupling approaches saturation and
significant X-ray heating begins. In most models, the absorption
trough is the strongest feature of the signal and its detection is one
of the main goals of the global 21-cm experiments; the large predicted
scatter in the location of this point should encourage observers to
search for the signal in as wide a frequency range as
possible. Measuring the redshift and depth of the absorption trough
alone would not help us to strongly constrain any single parameter,
but can be used to rule out some areas of the astrophysical parameter
space. Despite this complexity, we have shown that the depth of the
absorption trough of the 21-cm signal is strongly correlated with the
ratio between the Ly$\alpha$ intensity and the X-ray heating rate, as
given by a corresponding fitting formula.

The 21-cm signal from the low-$z$ maximum is typically expected to be
seen in emission, once heating approaches saturation (unless heating
occurs very late). This maximum is affected by both cosmic heating and
reionization, so in our models it is affected by both the total CMB
optical depth and the properties of X-ray sources, with some scatter
introduced by other parameters. For maxima that take place at late
times the signal is weaker since reionization is then more
advanced. We have fit simple functions to relations between (1) the
redshift and the brightness temperature of this point, (2) the heating
rate and the brightness temperature, and (3) the ratio of the heating
rate to the ionization production and the brightness temperature.
Therefore, measurement of the redshift and temperature at the emission
peak would give a self-consistency check and allow us to estimate both
the X-ray and ionizing intensity of sources.

Taken together, the correlations in Eqs.~(\ref{eq:JA}),
(\ref{eq:dJA}), (\ref{eq:minJdEPS}), (\ref{eq:lowzZ}), and
(\ref{eq:lowzZratio}), can be used to directly link future
measurements of the global 21-cm signal to astrophysical properties of
the high redshift Universe, in a mostly model-independent
way. Meanwhile, those in Eqs.~(\ref{eq:highz}) and (\ref{eq:lowzTZ})
can be used as consistency checks on the measurements (or on the
theory, depending on one's point of view). Some caution is
  advisable, as models like ours do not capture the full possible
  complexity of high-redshift astrophysics. For example, $f_*$ and the
  other efficiency parameters could vary with redshift, with the local
  density, or show a large scatter among halos. We expect the main
  effect of this to be that the correlations that we identified at
  each key turning point will measure the astrophysical parameters as
  averaged spatially and over time (out to an earlier time than that
  corresponding to a given key feature). Also, the scatter could
  increase in some correlations, particularly those that depend on
  multiple redshifts as in Figure~\ref{fig:Slope}. We plan to explore
  how more elaborate models affect our results.
Some of our conclusions are reminiscent of those found by \cite{Mirocha:2013}, 
now derived in the context of a wider array of astrophysical
models and more realistic simulations; a conclusion that is
particularly similar is that the high-$z$ maximum point reflects the Ly$\alpha$
intensity and its time derivative.
 
We have tried to cover as large a parameter space as possible, in
terms of astrophysical source formation, radiative efficiencies,
feedback effects, and the mean free path of ionizing photons. The goal
was to make our conclusions as robust as possible given current
uncertainties about high-redshift astrophysics. However, in the
results we have focused only on some of the parameters and a few
correlations, namely those that were cleanest and thus most useful.
We have explored many others that did not give a clearly useful
result, and we plan to continue such studies.

Current and future 21-cm observations, such as those mention in the
Introduction, are expected to soon begin to exclude realistic possible
realizations of the global 21-cm signal. We hope this is followed soon
afterwards with detections, which will probe currently mysterious
astrophysical processes at very high redshifts.

\section{Acknowledgments}
We thank Joe Lazio for a suggestion that led us to Figure~10.  R.B.\,
A.C.\, and M.L.\ acknowledge Israel Science Foundation grant 823/09
and the Ministry of Science and Technology, Israel. For R.B.\ and
A.C.\, this project/publication was made possible through the support
of a grant from the John Templeton Foundation. The opinions expressed
in this publication are those of the author(s) and do not necessarily
reflect the views of the John Templeton Foundation. R.B.'s work has
been partly done within the Labex Institut Lagrange de Paris (ILP,
reference ANR-10-LABX-63) part of the Idex SUPER, and received
financial state aid managed by the Agence Nationale de la Recherche,
as part of the programme Investissements d'avenir under the reference
ANR-11-IDEX-0004-02. R.B. also acknowledges a Leverhulme Trust
Visiting Professorship at the University of Oxford. This research was
supported in part by Perimeter Institute for Theoretical
Physics. Research at Perimeter Institute is supported by the
Government of Canada through the Department of Innovation, Science and
Economic Development Canada and by the Province of Ontario through the
Ministry of Research, Innovation and Science.




\appendix
\section{Cases List}
\label{appA}

\begin{table*}
	\begin{center}
	\begin{tabular}{lllllllllll}
		\hline \#  &                       & $f_*$ & $V_c$ [km/s] & $f_X$ & SED        & $\tau$ & LW  & Low-mass cutoff                & $\zeta$ & $R_{\rm mfp}$ [Mpc]\\
		\hline 1   & Filler                & 0.005 & 4.2   & 0.1   & Hard       & 0.066  & On  & Eq.~(\ref{eq:fstarGrad})  & 20      & 70 \\
		2   & Filler                & 0.005 & 4.2   & 0.1   & Hard       & 0.082  & On  & Eq.~(\ref{eq:fstarGrad})  & 32      & 70 \\
		3   & Filler                & 0.005 & 4.2   & 0.1   & Soft       & 0.066  & On  & Eq.~(\ref{eq:fstarGrad})  & 20      & 70 \\
		4   & Filler                & 0.005 & 4.2   & 0.1   & Soft       & 0.082  & On  & Eq.~(\ref{eq:fstarGrad})  & 32      & 70 \\
		5   & Filler                & 0.005 & 4.2   & 1     & Hard       & 0.066  & On  & Eq.~(\ref{eq:fstarGrad})  & 20      & 70 \\
		6   & Filler                & 0.005 & 4.2   & 1     & Hard       & 0.082  & On  & Eq.~(\ref{eq:fstarGrad})  & 32      & 70 \\
		7   & Filler                & 0.005 & 4.2   & 1     & Soft       & 0.066  & On  & Eq.~(\ref{eq:fstarGrad})  & 20      & 70 \\
		8   & Filler                & 0.005 & 4.2   & 1     & Soft       & 0.082  & On  & Eq.~(\ref{eq:fstarGrad})  & 32      & 70 \\
		9   & Filler                & 0.005 & 4.2   & 8     & Hard       & 0.066  & On  & Eq.~(\ref{eq:fstarGrad})  & 20      & 70 \\
		10  & Filler                & 0.005 & 4.2   & 8     & Hard       & 0.082  & On  & Eq.~(\ref{eq:fstarGrad})  & 32      & 70 \\
		11  & Filler                & 0.005 & 4.2   & 8     & Soft       & 0.066  & On  & Eq.~(\ref{eq:fstarGrad})  & 19      & 70 \\
		12  & Filler                & 0.005 & 4.2   & 8     & Soft       & 0.082  & On  & Eq.~(\ref{eq:fstarGrad})  & 31      & 70 \\
		13  & Filler                & 0.05  & 4.2   & 0.1   & Hard       & 0.066  & On  & Eq.~(\ref{eq:fstarGrad})  & 20      & 70 \\
		14  & Filler                & 0.05  & 4.2   & 0.1   & Hard       & 0.082  & On  & Eq.~(\ref{eq:fstarGrad})  & 36      & 70 \\
		15  & Filler                & 0.05  & 4.2   & 0.1   & Soft       & 0.066  & On  & Eq.~(\ref{eq:fstarGrad})  & 20      & 70 \\
		16  & Filler                & 0.05  & 4.2   & 0.1   & Soft       & 0.082  & On  & Eq.~(\ref{eq:fstarGrad})  & 36      & 70 \\
		17  & Filler                & 0.05  & 4.2   & 1     & Hard       & 0.066  & On  & Eq.~(\ref{eq:fstarGrad})  & 20      & 70 \\
		18  & Filler                & 0.05  & 4.2   & 1     & Hard       & 0.082  & On  & Eq.~(\ref{eq:fstarGrad})  & 36      & 70 \\
		19  & Filler                & 0.05  & 4.2   & 1     & Soft       & 0.066  & On  & Eq.~(\ref{eq:fstarGrad})  & 19      & 70 \\
		20  & Filler                & 0.05  & 4.2   & 1     & Soft       & 0.082  & On  & Eq.~(\ref{eq:fstarGrad})  & 35      & 70 \\
		21  & Filler                & 0.05  & 4.2   & 8     & Hard       & 0.066  & On  & Eq.~(\ref{eq:fstarGrad})  & 18      & 70 \\
		22  & Filler                & 0.05  & 4.2   & 8     & Hard       & 0.082  & On  & Eq.~(\ref{eq:fstarGrad})  & 34      & 70 \\
		23  & Filler                & 0.05  & 4.2   & 8     & Soft       & 0.066  & On  & Eq.~(\ref{eq:fstarGrad})  & 15      & 70 \\
		24  & Filler                & 0.05  & 4.2   & 8     & Soft       & 0.082  & On  & Eq.~(\ref{eq:fstarGrad})  & 31      & 70 \\
		25  & Filler                & 0.5   & 4.2   & 0.1   & Hard       & 0.066  & On  & Eq.~(\ref{eq:fstarGrad})  & 26      & 70 \\
		26  & Filler                & 0.5   & 4.2   & 0.1   & Hard       & 0.082  & On  & Eq.~(\ref{eq:fstarGrad})  & 51      & 70 \\
		27  & Filler                & 0.5   & 4.2   & 0.1   & Soft       & 0.066  & On  & Eq.~(\ref{eq:fstarGrad})  & 25      & 70 \\
		28  & Filler                & 0.5   & 4.2   & 0.1   & Soft       & 0.082  & On  & Eq.~(\ref{eq:fstarGrad})  & 51      & 70 \\
		29  & Filler                & 0.5   & 4.2   & 1     & Hard       & 0.066  & On  & Eq.~(\ref{eq:fstarGrad})  & 24      & 70 \\
		30  & Filler                & 0.5   & 4.2   & 1     & Hard       & 0.082  & On  & Eq.~(\ref{eq:fstarGrad})  & 50      & 70 \\
		31  & Filler                & 0.5   & 4.2   & 1     & Soft       & 0.066  & On  & Eq.~(\ref{eq:fstarGrad})  & 20      & 70 \\
		32  & Filler                & 0.5   & 4.2   & 1     & Soft       & 0.082  & On  & Eq.~(\ref{eq:fstarGrad})  & 44      & 70 \\
		33  & Filler                & 0.5   & 4.2   & 8     & Hard       & 0.066  & On  & Eq.~(\ref{eq:fstarGrad})  & 16      & 70 \\
		34  & Filler                & 0.5   & 4.2   & 8     & Hard       & 0.082  & On  & Eq.~(\ref{eq:fstarGrad})  & 39      & 70 \\
		35  & Filler                & 0.5   & 4.2   & 8     & Soft       & 0.066  & On  & Eq.~(\ref{eq:fstarGrad})  & 1.5     & 70 \\
		36  & Filler                & 0.5   & 4.2   & 8     & Soft       & 0.082  & On  & Eq.~(\ref{eq:fstarGrad})  & 19      & 70 \\
		37  & Filler                & 0.005 & 16.5  & 0.1   & Hard       & 0.066  & -   & -                              & 20      & 70 \\
		38  & Filler                & 0.005 & 16.5  & 0.1   & Hard       & 0.082  & -   & -                              & 37      & 70 \\
		39  & Filler                & 0.005 & 16.5  & 0.1   & Soft       & 0.066  & -   & -                              & 20      & 70 \\
		40  & Filler                & 0.005 & 16.5  & 0.1   & Soft       & 0.082  & -   & -                              & 37      & 70 \\
		41  & Filler                & 0.005 & 16.5  & 1     & Hard       & 0.066  & -   & -                              & 20      & 70 \\
		42  & Filler                & 0.005 & 16.5  & 1     & Hard       & 0.082  & -   & -                              & 37      & 70 \\
		43  & Filler                & 0.005 & 16.5  & 1     & Soft       & 0.066  & -   & -                              & 20      & 70 \\
		44  & Filler                & 0.005 & 16.5  & 1     & Soft       & 0.082  & -   & -                              & 36      & 70 \\
		45  & Filler                & 0.005 & 16.5  & 8     & Hard       & 0.066  & -   & -                              & 20      & 70 \\
		46  & Filler                & 0.005 & 16.5  & 8     & Hard       & 0.082  & -   & -                              & 36      & 70 \\
		47  & Filler                & 0.005 & 16.5  & 8     & Soft       & 0.066  & -   & -                              & 19      & 70 \\
		48  & Filler                & 0.005 & 16.5  & 8     & Soft       & 0.082  & -   & -                              & 36      & 70 \\
		49  & Filler                & 0.05  & 16.5  & 0.1   & Hard       & 0.066  & -   & -                              & 20      & 70 \\
		50  & Filler                & 0.05  & 16.5  & 0.1   & Hard       & 0.082  & -   & -                              & 37      & 70 \\
		51  & Filler                & 0.05  & 16.5  & 0.1   & Soft       & 0.066  & -   & -                              & 20      & 70 \\
		52  & Filler                & 0.05  & 16.5  & 0.1   & Soft       & 0.082  & -   & -                              & 36      & 70 \\
		53  & Standard              & 0.05  & 16.5  & 1     & Hard       & 0.066  & -   & -                              & 20      & 70 \\
		54  & Filler                & 0.05  & 16.5  & 1     & Hard       & 0.082  & -   & -                              & 36      & 70 \\
		55  & Filler                & 0.05  & 16.5  & 1     & Soft       & 0.066  & -   & -                              & 19      & 70 \\
		56  & Filler                & 0.05  & 16.5  & 1     & Soft       & 0.082  & -   & -                              & 36      & 70 \\
		57  & Filler                & 0.05  & 16.5  & 8     & Hard       & 0.066  & -   & -                              & 19      & 70 \\
		58  & Filler                & 0.05  & 16.5  & 8     & Hard       & 0.082  & -   & -                              & 35      & 70 \\
		59  & Filler                & 0.05  & 16.5  & 8     & Soft       & 0.066  & -   & -                              & 15      & 70 \\
		60  & Filler                & 0.05  & 16.5  & 8     & Soft       & 0.082  & -   & -                              & 31      & 70 \\
		61  & Filler                & 0.5   & 16.5  & 0.1   & Hard       & 0.066  & -   & -                              & 20      & 70 \\
		62  & Filler                & 0.5   & 16.5  & 0.1   & Hard       & 0.082  & -   & -                              & 36      & 70 \\
		63  & Filler                & 0.5   & 16.5  & 0.1   & Soft       & 0.066  & -   & -                              & 19      & 70 \\
		64  & Filler                & 0.5   & 16.5  & 0.1   & Soft       & 0.082  & -   & -                              & 36      & 70 \\
		65  & Filler                & 0.5   & 16.5  & 1     & Hard       & 0.066  & -   & -                              & 18      & 70 \\
		66  & Filler                & 0.5   & 16.5  & 1     & Hard       & 0.082  & -   & -                              & 35      & 70 \\
			\end{tabular}
		\end{center}
	\end{table*}
	\begin{table*}
		\begin{center}
			\begin{tabular}{lllllllllll}
				\hline \#  &                       & $f_*$ & $V_c$ [km/s]& $f_X$ & SED        & $\tau$ & LW  & Low-mass cutoff                & $\zeta$ & $R_{\rm mfp}$ [Mpc]\\
				\hline	
		67  & Filler                & 0.5   & 16.5  & 1     & Soft       & 0.066  & -   & -                              & 14      & 70 \\
		68  & Filler                & 0.5   & 16.5  & 1     & Soft       & 0.082  & -   & -                              & 30      & 70 \\
		69  & Filler                & 0.5   & 16.5  & 8     & Hard       & 0.066  & -   & -                              & 11.5    & 70 \\
		70  & Filler                & 0.5   & 16.5  & 8     & Hard       & 0.082  & -   & -                              & 27      & 70 \\
		71  & Filler                & 0.5   & 16.5  & 8     & Soft       & 0.066  & -   & -                              & 0.01    & 70 \\
		72  & Filler                & 0.5   & 16.5  & 8     & Soft       & 0.082  & -   & -                              & 9       & 70 \\
		73  & Filler                & 0.005 & 35.5  & 0.1   & Hard       & 0.082  & -   & -                              & 130     & 70 \\
		74  & Filler                & 0.005 & 35.5  & 0.1   & Soft       & 0.066  & -   & -                              & 53      & 70 \\
		75  & Filler                & 0.005 & 35.5  & 0.1   & Soft       & 0.082  & -   & -                              & 130     & 70 \\
		76  & Filler                & 0.005 & 35.5  & 1     & Hard       & 0.066  & -   & -                              & 53      & 70 \\
		77  & Filler                & 0.005 & 35.5  & 1     & Hard       & 0.082  & -   & -                              & 130     & 70 \\
		78  & Filler                & 0.005 & 35.5  & 1     & Soft       & 0.066  & -   & -                              & 53      & 70 \\
		79  & Filler                & 0.005 & 35.5  & 1     & Soft       & 0.082  & -   & -                              & 130     & 70 \\
		80  & Filler                & 0.005 & 35.5  & 8     & Hard       & 0.066  & -   & -                              & 53      & 70 \\
		81  & Filler                & 0.005 & 35.5  & 8     & Hard       & 0.082  & -   & -                              & 130     & 70 \\
		82  & Filler                & 0.005 & 35.5  & 8     & Soft       & 0.066  & -   & -                              & 53      & 70 \\
		83  & Filler                & 0.005 & 35.5  & 8     & Soft       & 0.082  & -   & -                              & 129     & 70 \\
		84  & Filler                & 0.05  & 35.5  & 0.1   & Hard       & 0.066  & -   & -                              & 53      & 70 \\
		85  & Filler                & 0.05  & 35.5  & 0.1   & Hard       & 0.082  & -   & -                              & 130     & 70 \\
		86  & Filler                & 0.05  & 35.5  & 0.1   & Soft       & 0.066  & -   & -                              & 53      & 70 \\
		87  & Filler                & 0.05  & 35.5  & 0.1   & Soft       & 0.082  & -   & -                              & 130     & 70 \\
		88  & Filler                & 0.05  & 35.5  & 1     & Hard       & 0.066  & -   & -                              & 53      & 70 \\
		89  & Filler                & 0.05  & 35.5  & 1     & Hard       & 0.082  & -   & -                              & 130     & 70 \\
		90  & Filler                & 0.05  & 35.5  & 1     & Soft       & 0.066  & -   & -                              & 52      & 70 \\
		91  & Filler                & 0.05  & 35.5  & 1     & Soft       & 0.082  & -   & -                              & 129     & 70 \\
		92  & Filler                & 0.05  & 35.5  & 8     & Hard       & 0.066  & -   & -                              & 52      & 70 \\
		93  & Filler                & 0.05  & 35.5  & 8     & Hard       & 0.082  & -   & -                              & 129     & 70 \\
		94  & Filler                & 0.05  & 35.5  & 8     & Soft       & 0.066  & -   & -                              & 48      & 70 \\
		95  & Filler                & 0.05  & 35.5  & 8     & Soft       & 0.082  & -   & -                              & 124     & 70 \\
		96  & Filler                & 0.5   & 35.5  & 0.1   & Hard       & 0.066  & -   & -                              & 53      & 70 \\
		97  & Filler                & 0.5   & 35.5  & 0.1   & Hard       & 0.082  & -   & -                              & 130     & 70 \\
		98  & Filler                & 0.5   & 35.5  & 0.1   & Soft       & 0.066  & -   & -                              & 52      & 70 \\
		99 & Filler                & 0.5   & 35.5  & 0.1   & Soft       & 0.082  & -   & -                              & 129     & 70 \\
		100 & Filler                & 0.5   & 35.5  & 1     & Hard       & 0.066  & -   & -                              & 52      & 70 \\
		101 & Filler                & 0.5   & 35.5  & 1     & Hard       & 0.082  & -   & -                              & 129     & 70 \\
		102 & Filler                & 0.5   & 35.5  & 1     & Soft       & 0.066  & -   & -                              & 47      & 70 \\
		103 & Filler                & 0.5   & 35.5  & 1     & Soft       & 0.082  & -   & -                              & 122     & 70 \\
		104 & Filler                & 0.5   & 35.5  & 8     & Hard       & 0.066  & -   & -                              & 43      & 70 \\
		105 & Filler                & 0.5   & 35.5  & 8     & Hard       & 0.082  & -   & -                              & 119     & 70 \\
		106 & Filler                & 0.5   & 35.5  & 8     & Soft       & 0.066  & -   & -                              & 22      & 70 \\
		107 & Filler                & 0.5   & 35.5  & 8     & Soft       & 0.082  & -   & -                              & 91      & 70 \\
		108 & Large                 & 0.5   & 4.2   & 10    & MQ         & 0.098  & Off & Eq.~(\ref{eq:fstarSharp}) & 27      & 70 \\
		109 & Large                 & 0.5   & 4.2   & 0.1   & Soft       & 0.098  & Off & Eq.~(\ref{eq:fstarSharp}) & 26      & 70 \\
		110 & Large                 & 0.5   & 4.2   & 0.1   & MQ         & 0.098  & Off & Eq.~(\ref{eq:fstarSharp}) & 28      & 70 \\
		111 & Large                 & 0.005 & 4.2   & 10    & Soft       & 0.098  & Off & Eq.~(\ref{eq:fstarSharp}) & 26      & 70 \\
		112 & Large                 & 0.005 & 4.2   & 10    & MQ         & 0.098  & Off & Eq.~(\ref{eq:fstarSharp}) & 27      & 70 \\
		113 & Large                 & 0.005 & 4.2   & 0.1   & Soft       & 0.098  & Off & Eq.~(\ref{eq:fstarSharp}) & 28      & 70 \\
		114 & Large                 & 0.005 & 4.2   & 0.1   & MQ         & 0.098  & Off & Eq.~(\ref{eq:fstarSharp}) & 28      & 70 \\
		115 & Large                 & 0.5   & 76.5  & 10    & Soft       & 0.066  & -   & -                              & 387     & 70 \\
		116 & Large                 & 0.5   & 76.5  & 10    & Soft       & 0.098  & -   & -                              & 6000    & 70 \\
		117 & Large                 & 0.5   & 76.5  & 10    & MQ         & 0.066  & -   & -                              & 450     & 70 \\
		118 & Large                 & 0.5   & 76.5  & 10    & MQ         & 0.098  & -   & -                              & 6060    & 70 \\
		119 & Large                 & 0.5   & 76.5  & 0.1   & Soft       & 0.066  & -   & -                              & 455     & 70 \\
		120 & Large                 & 0.5   & 76.5  & 0.1   & Soft       & 0.098  & -   & -                              & 6060    & 70 \\
		121 & Large                 & 0.5   & 76.5  & 0.1   & MQ         & 0.098  & -   & -                              & 6060    & 70 \\
		122 & Large                 & 0.016 & 76.5  & 10    & Soft       & 0.066  & -   & -                              & 455     & 70 \\
		123 & Large                 & 0.16 & 76.5  & 10    & Soft       & 0.098  & -   & -                              & 6060    & 70 \\
		124 & Large                 & 0.016 & 76.5  & 10    & MQ         & 0.066  & -   & -                              & 455     & 70 \\
		125 & Large                 & 0.16 & 76.5  & 10    & MQ         & 0.098  & -   & -                              & 6060    & 70 \\
		126 & Large                 & 0.16 & 76.5  & 0.1   & Soft       & 0.098  & -   & -                              & 6060    & 70 \\
		127 & Large                 & 0.16 & 76.5  & 0.1   & MQ         & 0.098  & -   & -                              & 6060    & 70 \\
			\end{tabular}
		\end{center}
	\end{table*}
	\begin{table*}
		\begin{center}

			\begin{tabular}{lllllllllll}
				\hline \#  &                       & $f_*$ & $V_c$ [km/s]& $f_X$ & SED        & $\tau$ & LW  & Low-mass cutoff                & $\zeta$ & $R_{\rm mfp}$ [Mpc] \\
				\hline
		128 & Small                 & 0.16 & 4.2   & 3.16  & Hard \& MQ & 0.066  & On  & Eq.~(\ref{eq:fstarGrad})  & 21      & 70 \\
		129 & Small                 & 0.16 & 4.2   & 3.16  & Hard \& MQ & 0.082  & On  & Eq.~(\ref{eq:fstarGrad})  & 41      & 70 \\
		130 & Small                 & 0.16 & 4.2   & 3.16  & Soft \& MQ & 0.066  & On  & Eq.~(\ref{eq:fstarGrad})  & 18      & 70 \\
		131 & Small                 & 0.16 & 4.2   & 3.16  & Soft \& MQ & 0.082  & On  & Eq.~(\ref{eq:fstarGrad})  & 38      & 70 \\
		132 & Small                 & 0.16 & 4.2   & 0.32  & Hard \& MQ & 0.066  & On  & Eq.~(\ref{eq:fstarGrad})  & 21      & 70 \\
		133 & Small                 & 0.16 & 4.2   & 0.32  & Hard \& MQ & 0.082  & On  & Eq.~(\ref{eq:fstarGrad})  & 42      & 70 \\
		134 & Small                 & 0.16 & 4.2   & 0.32  & Soft \& MQ & 0.066  & On  & Eq.~(\ref{eq:fstarGrad})  & 21      & 70 \\
		135 & Small                 & 0.16 & 4.2   & 0.32  & Soft \& MQ & 0.082  & On  & Eq.~(\ref{eq:fstarGrad})  & 42      & 70 \\
		136 & Small                 & 0.016 & 4.2   & 3.16  & Hard \& MQ & 0.066  & On  & Eq.~(\ref{eq:fstarGrad})  & 20      & 70 \\
		137 & Small                 & 0.016 & 4.2   & 3.16  & Hard \& MQ & 0.082  & On  & Eq.~(\ref{eq:fstarGrad})  & 33      & 70 \\
		138 & Small                 & 0.016 & 4.2   & 3.16  & Soft \& MQ & 0.066  & On  & Eq.~(\ref{eq:fstarGrad})  & 19      & 70 \\
		139 & Small                 & 0.016 & 4.2   & 3.16  & Soft \& MQ & 0.082  & On  & Eq.~(\ref{eq:fstarGrad})  & 33      & 70 \\
		140 & Small                 & 0.016 & 4.2   & 0.32  & Hard \& MQ & 0.066  & On  & Eq.~(\ref{eq:fstarGrad})  & 20      & 70 \\
		141 & Small                 & 0.016 & 4.2   & 0.32  & Hard \& MQ & 0.082  & On  & Eq.~(\ref{eq:fstarGrad})  & 33      & 70 \\
		142 & Small                 & 0.016 & 4.2   & 0.32  & Soft \& MQ & 0.066  & On  & Eq.~(\ref{eq:fstarGrad})  & 20      & 70 \\
		143 & Small                 & 0.016 & 4.2   & 0.32  & Soft \& MQ & 0.082  & On  & Eq.~(\ref{eq:fstarGrad})  & 33      & 70 \\
		144 & Small                 & 0.16 & 35.5  & 3.16  & Hard \& MQ & 0.066  & -   & -                              & 52      & 70 \\
		145 & Small                 & 0.16 & 35.5  & 3.16  & Hard \& MQ & 0.082  & -   & -                              & 129     & 70 \\
		146 & Small                 & 0.16 & 35.5  & 3.16  & Soft \& MQ & 0.066  & -   & -                              & 49      & 70 \\
		147 & Small                 & 0.16 & 35.5  & 3.16  & Soft \& MQ & 0.082  & -   & -                              & 126     & 70 \\
		148 & Small                 & 0.16 & 35.5  & 0.32  & Hard \& MQ & 0.066  & -   & -                              & 53      & 70 \\
		149 & Small                 & 0.16 & 35.5  & 0.32  & Hard \& MQ & 0.082  & -   & -                              & 130     & 70 \\
		150 & Small                 & 0.16 & 35.5  & 0.32  & Soft \& MQ & 0.066  & -   & -                              & 53      & 70 \\
		151 & Small                 & 0.16 & 35.5  & 0.32  & Soft \& MQ & 0.082  & -   & -                              & 130     & 70 \\
		152 & Small                 & 0.016 & 35.5  & 3.16  & Hard \& MQ & 0.066  & -   & -                              & 53      & 70 \\
		153 & Small                 & 0.016 & 35.5  & 3.16  & Hard \& MQ & 0.082  & -   & -                              & 130     & 70 \\
		154 & Small                 & 0.016 & 35.5  & 3.16  & Soft \& MQ & 0.066  & -   & -                              & 53      & 70 \\
		155 & Small                 & 0.016 & 35.5  & 3.16  & Soft \& MQ & 0.082  & -   & -                              & 130     & 70 \\
		156 & Small                 & 0.016 & 35.5  & 0.32  & Hard \& MQ & 0.066  & -   & -                              & 53      & 70 \\
		157 & Small                 & 0.016 & 35.5  & 0.32  & Hard \& MQ & 0.082  & -   & -                              & 130     & 70 \\
		158 & Small                 & 0.016 & 35.5  & 0.32  & Soft \& MQ & 0.066  & -   & -                              & 53      & 70 \\
		159 & Small                 & 0.016 & 35.5  & 0.32  & Soft \& MQ & 0.082  & -   & -                              & 130     & 70 \\
		160 & \citet{Fialkov:2016} & 0.05  & 16.5  & 1     & MQ         & 0.0738   & -   & -                              & 24      & 70 \\
		161 & \citet{Fialkov:2016} & 0.05  & 16.5  & 1     & MQ         & 0.0956  & -   & -                              & 57      & 70 \\
		162 & \citet{Fialkov:2016} & 0.05  & 16.5  & 10.8  & Hard       & 0.0756   & -   & -                              & 24      & 70 \\
		163 & \citet{Fialkov:2016} & 0.05  & 16.5  & 29.5  & Soft       & 0.0859   & -   & -                              & 24      & 70 \\
		164 & \citet{Fialkov:2016} & 0.05  & 16.5  & 11.4  & MQ         & 0.0747   & -   & -                              & 24      & 70 \\
		165 & \citet{Fialkov:2016} & 0.05  & 16.5  & 44.4  & Hard       & 0.0990  & -   & -                              & 57      & 70 \\
		166 & \citet{Fialkov:2016} & 0.05  & 16.5  & 102   & Soft       & 0.1111  & -   & -                              & 57      & 70 \\
		167 & \citet{Fialkov:2016} & 0.05  & 16.5  & 74.4  & MQ         & 0.0977  & -   & -                              & 57      & 70 \\
		168 & \citet{Fialkov:2016} & 0.05  & 16.5  & 0.01  & Hard       & 0.0739   & -   & -                              & 24      & 70 \\
		169 & \citet{Fialkov:2016} & 0.05  & 16.5  & 0.0023     & Soft       & 0.0746   & -   & -                              & 24      & 70 \\
		170 & \citet{Fialkov:2016} & 0.05  & 16.5  & 0     & Hard       & 0.0957  & -   & -                              & 57      & 70 \\
		171 & \citet{Fialkov:2016} & 0.05  & 35.5  & 1     & MQ         & 0.0597   & -   & -                              & 32      & 70 \\
		172 & \citet{Fialkov:2016} & 0.05  & 35.5  & 1     & MQ         & 0.0831  & -   & -                              & 112     & 70 \\
		173 & \citet{Fialkov:2016} & 0.05  & 35.5  & 14.7  & Hard       & 0.0609   & -   & -                              & 32      & 70 \\
		174 & \citet{Fialkov:2016} & 0.05  & 35.5  & 41.4  & Soft       & 0.0688   & -   & -                              & 32      & 70 \\
		175 & \citet{Fialkov:2016} & 0.05  & 35.5  & 12.1  & MQ         & 0.0606   & -   & -                              & 32      & 70 \\
		176 & \citet{Fialkov:2016} & 0.05  & 35.5  & 79.2  & Hard       & 0.0850  & -   & -                              & 112     & 70 \\
		177 & \citet{Fialkov:2016} & 0.05  & 35.5  & 188   & Soft       & 0.0934  & -   & -                              & 112     & 70 \\
		178 & \citet{Fialkov:2016} & 0.05  & 35.5  & 87.9  & MQ         & 0.0847  & -   & -                              & 112     & 70 \\
		179 & \citet{Fialkov:2016} & 0.05  & 35.5  & 0     & Hard       & 0.0831  & -   & -                              & 112     & 70 \\
		180 & \citet{Fialkov:2016} & 0.05  & 35.5  & 0.036  & Hard       & 0.0597   & -   & -                              & 32      & 70 \\
		181 & \citet{Fialkov:2016} & 0.05  & 35.5  & 0.0095  & Soft       & 0.0601   & -   & -                              & 32  & 70 \\
		182 & $R_{\rm mfp}$ & 0.005  & 35.5  & 0.1  & Hard       & 0.082   & -   & -                              & 125  & 20 \\
		183 & $R_{\rm mfp}$ & 0.005  & 35.5  & 0.1  & Soft       & 0.082   & -   & -                              & 125  & 20 \\
		184 & $R_{\rm mfp}$ & 0.005  & 35.5  & 1  & Hard       & 0.082   & -   & -                              & 125  & 20 \\
		185 & $R_{\rm mfp}$ & 0.05  & 35.5  & 0.1  & Hard       & 0.082   & -   & -                              & 125  & 20 \\
		186 & $R_{\rm mfp}$ & 0.5  & 76.5  & 0.1  & Hard       & 0.066   & -   & -                              & 389  & 20 \\
		187 & $R_{\rm mfp}$ & 0.5  & 76.5  & 0.1  & Hard       & 0.082   & -   & -                              & 1411  & 20 \\
		188 & $R_{\rm mfp}$ & 0.005  & 35.5  & 0.1  & Hard       & 0.082   & -   & -                              & 202  & 5 \\
		189 & $R_{\rm mfp}$ & 0.005  & 35.5  & 0.1  & Soft       & 0.082   & -   & -                              & 202  & 5 \\
		190 & $R_{\rm mfp}$ & 0.005  & 35.5  & 1  & Hard       & 0.082   & -   & -                              & 202  & 5 \\
		191 & $R_{\rm mfp}$ & 0.05  & 35.5  & 0.1  & Hard       & 0.082   & -   & -                              & 202  & 5 \\
		192 & $R_{\rm mfp}$ & 0.5  & 76.5  & 0.1  & Hard       & 0.066   & -   & -                              & 1172  & 5 \\
		193 & $R_{\rm mfp}$ & 0.5  & 76.5  & 0.1  & Hard       & 0.082   & -   & -                              & 6125  & 5 \\
		\hline   
	\end{tabular}
				\caption{List of the parameter sets
                                  used in this paper. }
				\label{Table:List}
\end{center}
\end{table*}

\end{document}